\documentclass[11pt,a4paper]{article}
\usepackage{jcappub}
\usepackage{rotating} 
\usepackage{graphicx,epsfig}
\usepackage{amsmath}
\usepackage {amssymb}
\usepackage{subfigure}

\usepackage{amssymb, amsmath, amsfonts, amsthm, graphicx}
\usepackage{pdflscape}
\def\bea{\begin{eqnarray}}
	\def\eea{\end{eqnarray}}
\def\parb{\pmb{\partial}}
\def\be{\begin{equation}}
	\def\ee{\end{equation}}
\def\e{\mathbf{e}}
\def\sp{\sigma_+}
\def\udot{\dot{u}}
\def\Ex{E_1{}^1}
\def\ex{e_1{}^1}
\def\ey{e_2{}^2}
\def\ez{e_3{}^3}
\def\R{{}^3\!R}
\def\S{{}^3\!S_+}
\def\y{\vartheta}
\def\z{\varphi}
\def\Udot{\dot{U}}

\def\sp{\sigma_+}
\def\Sp{\Sigma_+}
\def\R{{}^3\!R}      
\def\S{{}^3\!S_+} 
\def\calS{\mathcal{S}_+}
\def\D{D}

\begin{document}
	
	\title{Spherically symmetric Einstein-aether perfect fluid models}

	\author[a]{\sc Alan A. Coley}
	
	\author[b]{\sc Genly Leon}
	
	\author[c]{\sc Patrik Sandin}
	
	\author[a]{\sc Joey Latta}

	\affiliation[a]{Department of Mathematics and Statistics, Dalhousie University, Halifax, Nova Scotia, Canada  B3H 3J5}
	\affiliation[b]{Instituto de F\'{\i}sica, Pontificia Universidad  Cat\'olica de Valpara\'{\i}so, Casilla 4950, Valpara\'{\i}so, Chile}
	\affiliation[c]{Max-Planck-Institut f{\"u}r Gravitationsphysik (Albert-Einstein-Institut), Am M{\"u}hlenberg 1, D-14476 Potsdam, Germany.}

	\emailAdd{aac@mathstat.dal.ca}
	
	\emailAdd{genly.leon@ucv.cl}
		
	\emailAdd{patrik.sandin@aei.mpg.de}
	
	\emailAdd{lattaj@mathstat.dal.ca}

\abstract{We investigate spherically symmetric cosmological models
	in Einstein-aether theory
	with a tilted (non-comoving) perfect fluid source.
	We use a 1+3 frame formalism and adopt the
	comoving aether gauge
	to derive the evolution equations, which form a
	well-posed system of first order partial differential equations in two variables.
	We then introduce
	normalized variables.
	The formalism is particularly well-suited for numerical computations and the
	study of the qualitative properties of the models, 
	which are also solutions
	of Horava gravity. 
	We study the local stability
	of the equilibrium points of the resulting dynamical system
	corresponding to physically
	realistic inhomogeneous cosmological models 
	and astrophysical objects with 
	values for the  parameters which are consistent with current constraints.
	In particular, 
	we consider dust models in ($\beta-$) normalized variables
	and derive a reduced (closed) evolution system
	and we obtain the general evolution equations
	for the spatially homogeneous 
	Kantowski-Sachs models using appropriate  bounded
	normalized variables. We then
	analyse these models, with special emphasis on
	the future asymptotic behaviour for different values of the parameters.
	Finally, we investigate static models  for a mixture of a (necessarily non-tilted) perfect fluid with a barotropic equations of state
	and a scalar field. }

\keywords{Spherical symmetry, Einstein-aether theory, perfect fluid}

	
	\maketitle


\section{Introduction}

Since the vacuum in
quantum gravity may determine a preferred rest frame
at the microscopic level,
gravitational Lorentz violation has been studied within the framework of general relativity (GR),
where the background tensor field(s) breaking the symmetry
must be dynamical \cite{Jacobson}.
Einstein-aether theory \cite{DJ,Jacobson:2000xp} consists of GR  coupled, at second derivative
order, to a dynamical timelike unit vector field,
the aether. In this effective field theory
approach, the aether
vector field  and the metric tensor  together determine the
local spacetime structure.

The aether spontaneously breaks
Lorentz invariance by picking out a preferred frame at each point in spacetime while maintaining
local rotational symmetry (breaking only the boost sector of the Lorentz symmetry).
Since the aether is a
unit vector, it is everywhere non-zero in any solution,
including flat spacetime. A systematic construction of an
Einstein-aether gravity theory with a Lorentz violating dynamical field that preserves locality and
covariance in the presence of an additional `aether'
vector field  has been presented \cite{DJ,Jacobson:2000xp,CarrJ,kann,Zlosnik:2006zu}.

In the infra-red limit of (extended)
Horava gravity \cite{Horava} [a candidate ultra-violet
completion in the consistent non-projectable extension of Horava-Lifschitz gravity], the aether vector is assumed
to be hypersurface-orthogonal; hence every hypersurface-orthogonal
Einstein-aether solution is a Horava solution (most of the solutions studied).
The relationship between 
Einstein-aether theory and Horava gravity is further clarified in \cite{TJab13}, where it is shown how
Horava gravity can formally be obtained from Einstein-aether theory in the limit that the twist coupling
constant goes to infinity.

Cosmological models
in aether theories of gravity
are currently of interest. The impact of Lorentz violation on the inflationary scenario has been
explored \cite{CarrJ,kann,Zlosnik:2006zu} (also see the review
\cite{BlasLim}).  {\footnote{We note that in scalar field models in
which the dimensionless parameters of the models are not constant
(e.g., depend on the scalar field), it was found that inflationary
solutions are possible even in the absence of a scalar field potential
\cite{kann}.} } In particular, the primordial spectra of perturbations
generated by inflation in the presence of a timelike Lorentz-violating
vector field has been computed, and the amplitude of perturbation
spectra were found to be modified which, in general, leads to a
violation of the inflationary consistency relationship
\cite{EugeneLim}.

In particular, it is of importance to
study inhomogeneous cosmologies, in both GR and alternative gravitational theories,
partially motivated by current cosmological observations.
Measurements of anisotropies of the cosmic microwave background (CMB) from experiments including the WMAP
\cite{Bennett:2012zja} and Planck \cite{Ade:2013zuv} satellites, have provided strong support for the standard model of cosmology
with dark energy (and specifically a cosmological constant, $\Lambda$).  However the latest measurements are in
tension with local measurements of the Hubble expansion rate from supernovae Ia \cite{Riess:2011yx} and other cosmological observables
which point towards a lower growth rate of large-scale structure (LSS) 
(which may be evidence for deviations from the standard $\Lambda$CDM cosmological model).  
The possible observation by the BICEP2
experiment \cite{ade} of B-mode polarisation in the CMB in excess of the
signal due to lensing would indicate the first detection of gravitational waves, perhaps generated
during an inflationary era. 
In particular, there is a growing body of work on the imprints of gravitational waves on
large-scale structure. 

In this paper we will study spherically symmetric Einstein-aether models. 
We shall study perfect fluid matter models in general, and various
subcases in particular.
In a companion paper \cite{gangIIsf} we study spherically symmetric Einstein-aether scalar field
models with an exponential self-interaction potential.
Einstein-aether  models with an exponential potential were recently studied 
in \cite{Barrow:2012qy,Sandin,Alhulaimi:2013sha}.

We shall use the 1+3 frame formalism \cite{EU,WE,SSSS} to write down the
evolution equations for non-comoving perfect fluid spherically symmetric models and show they form a
well-posed system of first order partial differential equations (PDEs) in two variables. 
We adopt the so-called
comoving aether gauge (which implies a preferred foliation,
the only remaining freedom is coordinate time and space reparameterization).
We introduce normalized variables.
The formalism is particularly well-suited for numerical and qualitative analysis \cite{coleybook}.

In particular, we derive the governing equations
for an aether and a tilted perfect fluid
assuming that the acceleration $\udot$ is non-zero
and introduce  (so-called $\beta$-) normalized variables (some
of the technical details are relegated to the Appendix B). 
The evolution equations are presented in various different forms.
We also
rigorously derive the
evolution equations when  $\udot=0$.
We also consider the special subset  $\dot{U}=v=0$
(where $\dot{U}$ is the normalized acceleration, $v$ is the tilt, and we also assume   
the model parameters $c_\theta=0$ and $c_\sigma \neq 0$) 
and derive
the final reduced phase space equations in normalized variables.
We briefly review the  Friedmann-Lema\^{\i}tre-Robertson-Walker (FLRW) models in which the source must be of the form of a comoving perfect fluid (or vacuum)
and the aether must be comoving. We study in detail a number of special cases
of particular physical interest.

We first consider dust models, which are of particular interest at late times. 
We investigate a special dust model 
with $\dot{U}=0$ and $v = 0$ in normalized variables
(assuming
$c_{\sigma } \neq 0$) and derive a reduced (closed) evolution system.
The FLRW models in this special dust model correspond to an equilibrium point.
We are particularly interested in the future asymptotic behaviour of the models 
for different values of the parameters. 
We then consider the spatially homogeneous 
Kantowski-Sachs models using appropriate 
normalized variables (non-$\beta-$normalized  variables which are bounded), 
and obtain the general evolution equations. A full global dynamical analysis
of these models is possible. 
We then consider a special case  and
analyse the qualitative behaviour
for physically reasonable  values of the parameters at both early and late times.
Finally, we consider static models for a mixture of a (necessarily non-tilted) perfect fluid with a barotropic equations of state
and a scalar field,
which are also of physical importance (although
perhaps more from the astrophysical point of view than from the cosmological one).
A brief discussion of the physical conclusions is presented at the end.


\subsection{The models}
The evolution equations
follow from the field equations (FE) derived from the Einstein-aether action \cite{DJ,Jacobson:2000xp}.
In an Einstein-aether model there will
be additional terms in the FE which include (see the technical details  in the Section \ref{aetheory}):

\begin{itemize}

\item The effects on the geometry from the anisotropy and inhomogeneities (e.g., the curvature)
of the spherically symmetric models under consideration.

\item The Einstein FE are generalised by the contribution of an additional
stress tensor, $T^{\ae}_{ab}$, for the aether field which depends on the
dimensionless parameters of the aether model
(e.g., ``the $c_i$''). In GR, all of the $c_i = 0$.
To study the effects of matter, we could perhaps assume the
corresponding GR values (or close to them) in the first instance.

\item When the phenomenology of theories with a preferred
frame is studied, it is generally assumed that this frame
coincides, at least roughly, with the cosmological rest
frame defined by the Hubble expansion of the universe.
In particular, in an isotropic and spatially homogeneous Friedmann
universe
the aether field will be aligned with the
(natural preferred CMB rest frame) cosmic
frame and is thus related to the
expansion rate of the universe.
In principle, the preferred frame determined by the
aether can be different from (i.e., tilted with respect to)
the CMB rest frame in spherically symmetric models.
This adds additional
terms to the aether stress tensor $T^{\ae}_{ab}$, which can be characterized by a hyperbolic tilt angle,
$v(t)$, measuring the boost of the aether relative to the (perfect fluid) CMB rest frame
\cite{CarrJ,kann}. The tilt is expected to decay to the future in anisotropic but spatially 
homogeneous models \cite{tilt}.

\end{itemize}

\subsection{Spherical symmetry}

All spherically symmetric aether fields are hypersurface
orthogonal and, hence, all spherically symmetric solutions
of aether theory will also be solutions of the IR limit of
Horava gravity. The converse is not true in general, but it does
hold in spherical symmetry for solutions with a regular center \cite{TJab13}.

The $c_i$ are dimensionless constants in the model.
When spherical
symmetry is imposed the aether is hypersurface orthogonal,
and so it has vanishing twist. Thus it is possible to set $c_4$ to zero without loss of
generality \cite{Jacobson}.
After the parameter redefinition to eliminate $c_4$, one is left with a 3-
dimensional parameter space. The $c_i$ contribute to the effective Newtonian
gravitational constant $G$; so a renormalization of the  parameters in the
model can be then used to set $8 \pi G =1$ (i.e., another 
condition on the $c_i$ can effectively be specified).
The remaining parameters in the model can be
characterized by two non-trivial constant parameters.
The other constraints imposed on the $c_i$
have been summarized in \cite{Jacobson} (e.g.,  see
equations 43-46 in \cite{Barausse:2011pu}; also see Appendix A).
In GR $c_i = 0$.
We shall study the qualitative properties of models with 
values for the non-GR parameters which are consistent with current constraints.

Some of the models studied in this
paper involve a static metric coupled to a stationary aether. This
situation will be referred to here as ``stationary spherical
symmetry". This case will be treated separately later. An important special case occurs when the aether is
parallel to the Killing vector. We refer to this special case as
a ``static aether".
A spherically
symmetric static vacuum solution is known explicitly \cite{Eling:2006df}.

\subsection{Stars and black holes}

Spherically symmetric static and stationary  solutions are physical important.
Unlike GR, Einstein-aether theory has a spherically symmetric
mode, corresponding to radial tilting of the aether.
The
time-independent spherically symmetric solutions and  black
holes were studied in
\cite{Eling:2006df} and \cite{Eling:2006ec}, respectively, and surveyed in \cite{Jacobson},
and recently revisited for a more viable coupling parameter $c_1$ in
\cite{Eling:2006df} and
\cite{Barausse:2011pu}.
In general, within
this same parameter space, the dynamics of the
cosmological scale factor and perturbations differ
little from GR, and non-rotating neutron star and
black hole solutions are quite close to those of GR.
A thorough examination of the fully nonlinear solutions
has not been carried out to date.
A fully nonlinear energy positivity has, however, been established for spherically symmetric solutions at
a moment of time symmetry \cite{Garfinkle:2011iw}.

Let us discuss this in more detail. There is a three-parameter family of
spherically symmetric static vacuum solutions~\cite{Eling:2006df}.
In the
Einstein-aether theory the aether vector and its derivative provide
two additional degrees of freedom at each point.
If asymptotic flatness is imposed and the mass is
fixed, there remains a one-parameter
family (i.e., imposing asymptotic flatness reduces
this to a two parameter family~\cite{Eling:2003rd}), whereas
GR has the unique Schwarzschild solution
(Birkhoff's theorem).
In GR asymptotic flatness is a
consequence of the vacuum field equations without any tuning of
initial data, so the one-parameter family of local (Schwarzschild)
solutions is automatically asymptotically flat.
The radial tilt of the aether provides another
local degree of freedom in aether theory, so spherical solutions need
not be time-independent (even when restricting to
stationary spherically symmetric aether theory). Not only are spherical solutions not generally
static, but even if we restrict to static, spherical solutions,
they are not necessarily asymptotically flat.
It was
shown in~\cite{Jacobson:2000xp} that the Reissner-Nordstrom metric
in a spherically symmetric static gauge with fixed norm is a
solution, although
this is not the only solution in that special case~\cite{Eling:2006df}.

Requiring that the aether be aligned with the
timelike Killing field restricts the  static aether solution to one parameter 
(the single parameter  $c_{14}$, essentially the total
mass \cite{Eling:2006df}).
Thus the solution outside a static star is the unique
vacuum solution for a given mass in the  static aether
case~\cite{Eling:2006df}, and is asymptotically flat. In
\cite{Seifert:2007fr} it was found that this
static ``wormhole'' aether solution is generally stable to linear
perturbations under the same conditions as for flat
spacetime. In the pure GR limit ($c_1=0$),
we have just the Schwarzschild solution.
For small values of $r$,
the solutions can behave quite differently from the Schwarzschild solution.
More recently,
an analytic static spherically symmetric vacuum solution in the Einstein-aether theory was presented  
(demonstrated numerically) by use of the
Euler-Lagrange equations \cite{Gao:2013im}.

Unlike the singular wormhole, the static solutions have a regular origin \cite{Eling:2006df}.
It is known that pure aether stars do
not exist; i.e., there are no asymptotically flat self-gravitating
aether solutions with a regular origin \cite{Eling:2006df}. 
It has been shown that in the presence of a perfect fluid,
regular asymptotically flat star solutions exist and are
parameterized (for a given equation of state) by the central
pressure (see also \cite{Eling:2007xh}).

For black holes the aether cannot be aligned with
the Killing vector, since the latter is not timelike
on and inside the horizon. Instead, the aether  is at
rest at spatial infinity and flows inward at finite
radii. The condition of regularity (at the spin-0
horizon) selects a unique solution from the
one-parameter family of spherical stationary solutions
for a given mass~\cite{Eling:2006ec,Eling:2006df}.
Such black holes are
rather close to Schwarzschild outside the horizon for
a wide range of couplings. Inside the horizon the
solutions differ more (but typically no more than a few percent), and like the Schwarzschild solution they
contain a spacelike singularity.

More recently, static spherically symmetric, asymptotically flat, regular
(non-rotating) black-hole solutions in Einstein-aether theory have been studied 
(numerically) \cite{Barausse:2011pu},
generalizing previous results. It has been found that spherical black-hole
solutions formed by gravitational collapse exist for all viable
parameter values of the theory and a notion of black hole thus
persists. Indeed, static spherically symmetric 
solutions in Lorentz-violating theories,
in which the causal structure of gravity is greatly modified,
still possess a special hypersurface, called a ``Universal horizon'',
that acts as a genuine absolute causal boundary because it traps all
excitations, even those which could be traveling at arbitrarily high
propagation speeds \cite{Barausse:2011pu}.  The Universal horizon
satisfies a first law of black-hole mechanics \cite{Berglund1}, and
evidence has been found that Hawking radiation is associated with the
Universal horizon \cite{Berglund2,Cropp}.

Finally, it would be
of interest to determine the structure of rotating solutions;
rapidly rotating black holes,
unlike the non-rotating ones, might turn out to be
very different from the Kerr metrics of GR.


\section{Spherically symmetric Einstein-aether Models}

We shall use the 1+3 frame formalism \cite{EU,WE} to write down the
evolution equations for spherically symmetric models as a
well-posed system of first order PDEs in two variables.
The formalism is particularly well-suited for studying
perfect fluid spherically symmetric
models \cite{SSSS}, and especially
for numerical and qualitative analysis \cite{coleybook}. 
We follow a similar approach to that in the resource paper \cite{Coley:2008qd}
(wherein all relevant quantitites are explicitly defined).

\subsection{Restrictions on the kinematic and auxiliary variables:}

The metric is:
\be
\label{metric}
ds^2 = - N^2 dt^2 + (\ex)^{-2} dx^2  + (\ey)^{-2} (d\y^2 + \sin^2 \y  d\z^2).
\ee
The Killing vector fields (KVF) are given by \cite{kramer}:
\be 
\partial_\z,\quad \cos \z \ \partial_\y - \sin \z \cot \y \ \partial_\z,\quad  \sin \z \ \partial_\y + \cos \z \cot \y \ \partial_\z.
\ee
The frame vectors in coordinate form are:
\be
    \e_0 = N^{-1} \partial_t
    ,\quad
    \e_1 = \ex \partial_x
        ,\quad
        \e_2 = \ey \partial_\y
        ,\quad
        \e_3 = \ez \partial_\z,
\ee
where $\ez = \ey / \sin \y$. $N$, $\ex$ and $\ey$ are functions of $t$
and $x$.

This leads to the following restrictions on the kinematic variables:
\be
    \sigma_{\alpha\beta} = \text{diag}(-2\sp,\sp,\sp),\quad
    \omega_{\alpha\beta} =0,\quad
    \udot_\alpha =(\udot_1,0,0),
\ee
where \be {\dot u_\alpha} = {u^\beta}{\nabla _\beta}{u_\alpha}; \ee
\be
    \udot_1 = \e_1 \ln N;
\ee
on the spatial commutation functions:
\be
    a_\alpha = (a_1, a_2, 0),\quad
    n_{\alpha\beta} = \left( \begin{array}{ccc}
            0 & 0 & n_{13}  \\
            0 & 0 & 0   \\
            n_{13} & 0 & 0 \end{array} \right),
\ee
where

\be
    a_1 = \e_1 \ln\ey,\quad
    a_2 = n_{13} = - \frac12 \ey \cot \y;
\ee
and on the matter components:
\be
    q_\alpha = (q_1,0,0),\quad
    \pi_{\alpha\beta} = \text{diag}(-2\pi_+,\pi_+,\pi_+).
\ee
The frame rotation $\Omega_{\alpha\beta}$ is also zero.

Furthermore, $n_{13}$ only appears in the equations together with
$\e_2 n_{13}$ in the form of the Gauss curvature of the spheres
\be
    {}^2\!K := 2(\e_2 - 2 n_{13}) n_{13},
\ee
which simplifies to
\be
    {}^2\!K = (\ey)^2.
\ee
Thus the dependence on $\y$ is hidden in the equations. We will
also use ${}^2\!K$ in place of $\ey$.

To simplify notation, we will write
\[
    {}^2\!K,\ \udot_1,\ a_1
\]
as
\[
    K,\ \udot,\ a.
\]
To summarize, the essential variables are
\be
    N, \ex,\ K,\ \theta,\ \sigma_+,\ a,\ \udot, \ \mu,\ q_1,\ p,\ \pi_+,
\ee
where $N$ is the lapse function, $\ex$ is the non null component of the frame vector $\e_1$, $K$ is the Gauss curvature of the spheres, $\theta$ is the (volume) rate of expansion scalar, $\sigma_+$ is related to the magnitude of the rate of shear tensor (a measure of the anisotropies present in the model), $a$ is the radial component of the object (spatial commutation function) $a_\alpha$, $\udot$ is the acceleration, $\mu$ denotes the total energy density scalar, $q_1$ is a component of the total energy current density vector, $p$ is the total isotropic  pressure scalar and $\pi_+$ is related to the magnitude of the total anisotropic pressure tensor \cite{EU,WE}.

In the
case of spherical symmetry in
Einstein-aether theory one must be careful in choosing
the  gauge  (an additional gauge condition).  
 Normally, in GR, spherically symmetric coordinates are chosen so that the metric is simplified (e.g.,  a
choice for $N$) or so that the fluid is comoving.  Here we 
chose the aether vector field to be aligned with the timelike frame vector $\e_0$
(the comoving aether gauge, 
and hence in general $N(t,x)$ cannot be simplified any further). This may make comparisons with
GR difficult in some special cases.
Our formulation is perhaps better suited for fluids/matter and cosmology, although the 
static case is not  necessarily aligned (see later).

We note that the tilt is defined relative to matter; one important question is to investigate whether this tilt decays
to the future.


\subsection{Einstein-aether theory}
\label{aetheory}

The action for Einstein-aether theory is the most general generally
covariant functional of the spacetime metric $g_{ab}$ and aether
field $u^a$ involving no more than two derivatives (not including
total derivatives) \cite{Jacobson,Garfinkle:2011iw}.
The action is \cite{Jacobson,Carroll:2004ai}:
\begin{equation}
S=\int d^{4}x\sqrt{-g}\left[  \frac{1}{2}R - K^{a b}{}_{c d
}\nabla_{a}u^{c}\nabla_{b}u^{d} + \lambda\left(  u^{c}u_{c
} + 1\right) + \mathcal{{L}}_m  \right]  ,\label{action}
\end{equation}
where
\begin{equation}
K^{a b}{}_{c d}\equiv c_{1}g^{a b}g_{c d} + c_{2}
\delta_{c}^{a}\delta_{d}^{b} + c_{3}\delta_{d}^{a}
\delta_{c}^{b} + c_{4}u^{a}u^{b}g_{c d}.
\end{equation}
The action (\ref{action}) contains an Einstein-Hilbert term for the metric, a kinetic term
for the aether with four dimensionless coefficients $c_{i}$, and $\lambda$
is a Lagrange multiplier enforcing the time-like  constraint on
the aether. \footnote{We set the vector norm to unity in order to obtain a unit time-like aether. 
Comparing with \cite{Garfinkle:2007bk,Garfinkle:2011iw}, the tensor  $K^{a b}_{m n}$ was rescaled by a factor of 2, 
$c_4$ was taken with the opposite sign, and $J_{a b}$ was redefined taking the 
opposite sign (i.e., the constant $c$'s here and $\lambda$ have been rescaled 
by a factor of 2).}
The convention used in this paper for the metric signature
is $({-}{+}{+}{+})$ and the units are chosen so that the speed of
light defined by the metric $g_{ab}$ is unity and $\kappa^2\equiv 8\pi G=1.$
The field equations from varying (\ref{action}) with respect to
$g^{ab}$, $u^a$, and $\lambda$ are given, respectively, by \cite{Garfinkle:2007bk}:
\bea
{G_{ab}} &=& {T^{TOT}_{ab}}
\label{EFE2}
\\
\lambda {u_b} &=& {\nabla _a} {{J^a}_b}+ c_4 \udot_a \nabla_b u^a
\label{evolveu}
\\
{u^a}{u_a} &=& -1.\label{unit}
 \eea 
Here $G_{ab}$ is the Einstein
tensor of the metric $g_{ab}$.  ${T^{TOT}_{ab}}$ 
is the {\em{total}} energy momentum tensor, ${T^{TOT}_{ab}}=T^{\ae}_{ab}+T^{mat}_{ab}$,
where $T^{mat}_{ab}$ is the total contribution from all matter sources. We shall omit
$T^{mat}_{ab}$ for the moment (and add in later for perfect fluid and scalar field
sources), and so we begin with the {\em{vacuum}} case ($\mathcal{{L}}_m=0$)
first with a non-trivial
aether stress-energy $T^{\ae}_{ab}$
(which we will refer to as ``pure" Einstein-aether 
theory which is a theory of the spacetime metric $g_{a b}$ and a
vector field (the ``aether") $u^{a}$).

The quantities ${J^a}_b,\; {\udot_a}$
and the aether stress-energy $T^{\ae}_{ab}$ are given by
\begin{subequations}
\begin{align} {{J^a}_m} & =
-{{K^{ab}}_{mn}}{\nabla_b}{u^n}
\label{J}\\
{\dot u_a} &= {u^b}{\nabla _b}{u_a}
\label{a}\\
\nonumber
{T^{\ae}_{ab}} &= 2c_{1}(\nabla_{a}u^{c}\nabla_{b}u_{c}-
\nabla^{c}u_{a}\nabla_{c}u_{b})\nonumber\\
&  - 2[\nabla_{c}(u_{(a} J^{c}{}_{b)}) + \nabla_{c}(u^{c
}J_{(a b)}) - \nabla_{c}(u_{(a}J_{b)}{}^{c})] -2 c_4 \udot_a \udot_b +\nonumber\\
&  + 2\lambda u_a u_b + g_{a b}\mathcal{L}_{u} \label{aestress}
\end{align}
\end{subequations}
where
\begin{equation}\label{aeLagrangian}
\mathcal{L}_{u} \equiv -K^{a b}{}_{c d}\nabla_{a}u^{c
}\nabla_{b}u^{d},
\end{equation} is the Einstein-aether Lagrangian \cite{Jacobson:2004ts}.

Taking the contraction of \eqref{evolveu} with $u^b$ and with the induced metric $h^{b c}:= g^{b c}+u^b u^c$ we obtain the equations
\begin{subequations}
\label{aether_eqs}
\begin{align}
\label{definition:lambda}
&\lambda = - u^b \nabla_a J^a_b-c_4 \udot_a \udot^a,\\
\label{restriction_aether}
& 0 = h^{b c}\nabla_a J^a_b + c_4 h^{b c} \udot_a \nabla_b u^a.
\end{align}
\end{subequations}
We shall use the equation \eqref{definition:lambda} as a definition for the Lagrange multiplier, whereas the second  equation \eqref{restriction_aether} leads to a set of restrictions that the aether vector must satisfy.  

The Einstein FE, Jacobi identities and contracted Bianchi identities gives a system of partial differential equations 
on the frame and commutator functions, while \eqref{aestress} defines the components of the energy momentum tensor and 
\eqref{restriction_aether} gives one extra equation for the aether. We choose a gauge in which the aether is aligned with $\e_0$,
the comoving aether temporal gauge (all that then remains is the time and space reparameterization 
freedom):~\footnote{Note that some degenerate cases, including the static case below, 
may not be easily included in this approach.}
\begin{subequations}
\begin{align}\label{evol_eq_vacuum}
& \e_0 (\ex) = - \tfrac{1}{3}(\theta - 6\sigma_+) \ex,
\\
& \e_0 (K) = - \tfrac{2}{3}(\theta + 3\sigma_+)K,
\\
& \e_0 (\theta) -  \e_1(\udot) = - \tfrac{1}{3}\theta^2 - 6\sigma_+^2 +(\udot - 2a)\udot - \frac{1}{2}(\mu + 3p),  
\\
& \e_0(\sigma_+) - \frac{1}{3} \e_1(\udot - a) = - \theta \sigma_+ - \frac{1}{3}(a + \udot)\udot - \frac{1}{3}K + \pi_+,
\\ 
& \e_0(a) = -\tfrac{1}{3}(\theta + 3\sigma_+)(a + \udot) - \tfrac{1}{2}q_1,
\\
& \e_0(\mu) + \e_1(q_1)= - \theta(\mu + p) + 2(a - \udot)q_1 - 6\sigma_+\pi_+ ,
\\
& \e_0(q_1) + \e_1(p) = -\tfrac{2}{3}(2\theta - 3\sigma_+)q_1 - 2(3a - \udot - \e_1)\pi_+ - \udot(\mu + p),
\end{align}
\end{subequations}
Constraints:
\begin{subequations}
\begin{align}
&  \e_1(\ln N) = \udot, \label{udot}
\\ 
& \e_1(\ln K) = 2a,
\\
& \mu = 3H^2 - 3\sigma_+^2 + K - 3a^2 + 2\e_1(a),
\\
& q_1 = - 6a\sigma_+ + \tfrac{2}{3}\e_1(\theta + 3\sigma_+), \label{q_constr}
\end{align}
\end{subequations}
where {\footnote{Here, for example,  
$\mu \equiv \mu^{tot}$ is the {\em{total}} energy density.  }}
$(\mu,\ p,\ q_1,\ \pi_+) = (\mu_{\ae},\ p_{\ae},\ q_{\ae},\ \pi_{\ae})$ can be computed from \eqref{aestress}:
\begin{subequations}
\begin{align}
& \mu_{\ae} = (c_1 - c_4)(2\e_1 - 4a + \udot)\udot - \tfrac{1}{3}(c_1 + 3c_2 + c_3)\theta^2 - 6(c_1 + c_3)\sigma_+^2,\label{aetherEnergy} \\
& p_{\ae} = \tfrac{1}{3}(c_1 + 3c_2 + c_3)(2\e_0 + \theta)\theta - 6(c_1 + c_3)\sigma_+^2 + \tfrac{1}{3}(c_1 - c_4)\udot^2, \\
& q_{\ae} = - \tfrac{4}{3}(c_1 - c_4)(\theta + 3\sigma_+)\udot - 2(c_1 - c_4)\e_0(\udot),\label{aetherHeatFlow}\\
& \pi_{\ae} = \tfrac{2}{3}(c_1 - c_4)\udot^2 + 2(c_1 + c_3)(\e_0 + \theta)\sigma_+.\label{aetherStress}
\end{align}
\end{subequations}

The aether equation \eqref{restriction_aether} becomes (and is true regardless of whether  $\udot$ is zero or not)
\begin{align}
\label{restriction:3}
(c_1 - c_4)\e_0(\udot) &= -\tfrac{2}{3}(c_1 - c_4)(\theta + 3\sigma_+)\udot + 6(c_1 + c_3)a\sigma_+ \nonumber \\ & + \tfrac{1}{3}(c_1 + 3c_2 + c_3)\e_1(\theta) - 2(c_1 + c_3)\e_1(\sigma_+).
\end{align}

To simplify these expressions it is convenient to make a reparameterization of the aether
parameters, analogous to the one given in~\cite{TJab13}:
\begin{displaymath}
c_\theta = c_2 + (c_1 + c_3)/3,\ c_\sigma = c_1 + c_3,\ c_\omega = c_1 - c_3,\ c_a = c_4 - c_1,
\end{displaymath}
where the new parameters correspond to terms in the Lagrangian relating to expansion,
shear, acceleration and twist of the aether.  Since the spherically symmetric models are
hypersurface orthogonal the aether field has vanishing twist and is therefore independent
of the twist parameter $c_\omega$ 
(the coupling $c_1-c_3$ does not occur in
the field equations (only $c_1+c_3$ does) \cite{TJab13}; this is equivalent to
being able to set $c_4=0$  \cite{Jacobson}).

A second condition on the $c_i$ can effectively be specified
by a renormalization of the Newtonian
gravitational constant $G$.
The remaining parameters in the model can therefore be
characterized by two non-trivial constant parameters. 
The other constraints imposed on the $c_i$
have been summarized in \cite{Jacobson}.

It may be useful later to define
$c^2 \equiv 1 - 2c_\sigma \leq 1$. 
In particular, some special cases of interest are (see the Appendix A):
{\bf{case A: $c_\sigma=\frac{1}{2}(1- c^2)  \geq 0, 
c_a =  -\frac{d}{(1+d)} c_\sigma \leq 0, c_\theta=0$: }}
{\bf{case B(ii): $ c_\sigma=\frac{1}{2}(1- c^2)  \geq 0,  c_a=-\frac{1}{2}(1- c^2),  
c_\theta=0$: }}
{\bf{case C: $c_\sigma=\frac{1}{2}(1- c^2)  \geq 0, 
c_\theta = -\frac{1}{3}(1- c^2) \leq 0, [c_a = 0]$ }}


The Lagrangian \eqref{aeLagrangian} becomes

\be
\label{eq:25}
\mathcal{L}_{u} = - \left(c_a \udot^2 + c_\theta \theta^2 + 6 c_\sigma \sigma_+^2\right)
\ee
and the aether energy components $(\mu,\ p,\ q_1,\ \pi_+) = (\mu_{\ae},\ p_{\ae},\ q_{\ae},\ \pi_{\ae})$ become
\begin{subequations}
\begin{align}
\mu_{\ae} &=  -c_\theta \theta^2 - 6c_\sigma \sigma_+^2 - c_a (\udot + 2\e_1 - 4a)\udot, \label{aetherEnergy_parAlt}
\\
p_{\ae} &=  -\tfrac{1}{3}c_a\udot^2 - 6c_{\sigma} \sigma_+^2 + c_{\theta}(2\e_0 + \theta)\theta,
\\
q_{\ae} &=  \tfrac{4}{3}c_a(\theta + 3\sigma_+)\udot + 2c_a\e_0(\udot),
\label{aetherHeatFlow_parAlt}
\\
\pi_{\ae} &=  - \tfrac{2}{3}c_a \udot^2 + 2c_{\sigma} (\e_0 + \theta)\sigma_+. \label{aetherStress_parAlt}
\end{align}
\end{subequations}

and the aether equation \eqref{restriction_aether} reads
\begin{align}
\label{aetherEq_parAlt}
& c_a\e_0(\udot) = -\tfrac{2}{3}c_a(\theta + 3\sigma_+)\udot - 6c_\sigma a \sigma_+ - \e_1(c_\theta \theta - 2c_\sigma \sigma_+).
\end{align}

Combining all of the above equations, and assuming $\udot\neq 0$  
(the special case $\udot = 0$ will be dealt with later), we obtain 
\begin{subequations}
\begin{align}
\label{synch_vacuum}
& \e_0 (\ex) = - \tfrac{1}{3}(\theta - 6\sigma_+) \ex,
\\
\label{evol_K}
& \e_0 (K) = - \tfrac{2}{3}(\theta + 3\sigma_+)K,
\\
\label{dudot_vacuum}
& \e_0(\udot) - \frac{\left(3 c_{\theta } + 2 c_{\sigma } \right) \e_1(\theta)}{3c_a \left(2 c_{\sigma }-1\right)}= -\tfrac{2}{3} \udot (\theta + 3\sigma_+), \\ 
& \e_0(\theta) - \frac{\left(c_a + 1 \right) \e_1(\udot)}{(3 c_{\theta } + 1)} = - \tfrac{1}{3}\theta^2 + \frac{\left(c_a + 1\right)\udot^2}{(3 c_{\theta } + 1)} - \frac{2 \left(c_a + 1\right) a \udot}{(3 c_{\theta } + 1)} + \frac{6 \left(2 c_{\sigma } - 1\right) \sigma_+^2}{(3 c_{\theta } + 1)}, \\		
& \e_0(\sigma_+) - \frac{\left(c_a + 1\right) \e_1(\udot)}{3 \left(2 c_{\sigma } - 1\right)} = - \theta\sigma_+ + \tfrac{1}{2} \sigma_+^2 + \frac{\left(3 c_{\theta } + 1\right) \theta^2}{18 \left(2 c_{\sigma } - 1\right)} + \frac{\left(1 - 2
c_a \right) a \udot}{3 \left(2 c_{\sigma }-1\right)} + \nonumber \\ & + \frac{\left(5 c_a + 2\right) \udot^2}{6 \left(2 c_{\sigma} - 1\right)} - \frac{a^2}{2 \left(2 c_{\sigma } - 1\right)} + \frac{K}{2 \left(2 c_{\sigma } - 1\right)}, \\
& \e_0(a) + \frac{\left(3 c_{\theta } + 2 c_{\sigma }\right) \e_1(\theta)}{6 c_{\sigma } - 3} = - \tfrac{1}{3}(a + \udot)(\theta + 3\sigma_+).
\end{align}
\end{subequations}
Constraints:
\begin{subequations}
\begin{align}
&  \e_1(\ln N) = \udot, \label{udot2}
\\ 
\label{constraint_K}
& \e_1(\ln K) = 2a,
\\
\label{eliminate_spatial}
& \e_1(a) + c_a \e_1(\udot) = -\tfrac{1}{6}\left(3 c_{\theta } + 1\right) \theta^2 - \frac{3}{2} \left(2 c_{\sigma } - 1\right) \sigma_{+}^2 - \frac{K}{2} + \frac{3 a^2}{2} + 2 c_a a\udot - \frac{c_a \udot^2}{2},
\\\label{constraint_sigma_H}
& \e_1(\sigma_+) - \frac{\left(3 c_{\theta } + 1\right) \e_1(\theta)}{6 c_{\sigma } - 3} = 3 a \sigma_+.
\end{align}
\end{subequations}
Commutator: 
\begin{equation}\label{comm}
[\e_0,\e_1]=\udot \e_0-\tfrac{1}{3}(\theta-6\sigma_+)\e_1.
\end{equation}

{\bf{Integrability conditions}}.
In the Einstein-aether analysis, one of the ``field equations'' is the spatial projection 
(with the induced
metric $h^{b c}$) of the equation
obtained by the contraction of the velocity variation of the action.  In many cases this equation does not involve the
appropriate time derivatives, and hence this equation is not an
evolution equation, but rather it is a constraint \cite{TJab11}. In the spherically symmetric case here
it can be shown that the constraint is conserved and is
compatible with all of  the other (evolution) equations.

\section{Aether and a tilted perfect fluid}

The energy momentum-tensor for the matter field is
  \be
 {T^{m}_{ab}}\equiv -2\frac{\delta \mathcal{L}_m}{\delta g^{a b}}+\mathcal{L}_m g_{a b}= \hat{\mu} u_a u_b + \hat{p} ( g_{ab} + u_a u_b).
 \ee
with $\hat{p}$ to be specified.
 In general, the 4-velocity vector $\mathbf{u}$ of the perfect fluid is not
 aligned with the vector $\e_0$ of a chosen temporal gauge. In spherically
 symmetric models, $\mathbf{u}$ is allowed to be of the form
\be
     \mathbf{u} = \Gamma(\e_0 + v \e_1),\quad
     \Gamma = (1-v^2)^{-\frac12},
\ee
where $v$ is the tilt parameter. We choose a linear equation of state for 
the perfect fluid:
\be
 \label{linear_eos}
     \hat{p} = (\gamma-1) \hat{\mu},
\ee
where $\gamma$ is a constant satisfying $1 \leq \gamma < 2$.
 Then we obtain for the tilted fluid:
\begin{subequations}
 \begin{align}
     \mu &= \frac{G_+}{1-v^2} \hat{\mu}
 \\
     p &= \frac{(\gamma-1)(1-v^2) + \frac13\gamma v^2}{1-v^2} \hat{\mu}
 \\
     q_1 &= \frac{\gamma \hat{\mu}}{1-v^2} v
 \\
     \pi_+ &= - \frac13 \frac{\gamma \hat{\mu}}{1-v^2} v^2,
 \end{align}
\end{subequations}
 where $G_\pm = 1 \pm (\gamma-1)v^2$.
 Thus (the total) $\mu$, $p$, $q_1$ and $\pi_+$ are given in terms of $\hat{\mu}$ and $v$.
These are then substituted into the evolution and constraint equations.

Assuming $\udot\neq 0$ (and under the general conditions
that $2c_\sigma -1 \neq 0, 3c_\theta -1 \neq 0$) we obtain:

\begin{subequations}
\begin{align}\label{synch}
    & \e_0 (\ex) = - \tfrac13 (\theta- 6\sigma_+) \ex,
\\
& \e_0 (K) = - \tfrac23 (\theta + 3 \sigma_+)K,
\\
 \label{dudot}
 & \e_0(\udot)-\frac{\left(3 c_{\theta }+2 c_{\sigma }\right) \e_1(\theta)}{3 c_a \left(2
   c_{\sigma }-1\right)}= -\tfrac23  \udot (\theta +3\sigma_+) +\frac{\gamma  c_{\sigma } \hat{\mu} v}{c_a \left(2 c_{\sigma }-1\right) (1-v^2)}, \\ 
	&\e_0(\theta)-\frac{\left(c_a+1\right) \e_1(\udot)}{(3 c_{\theta }+1)}=-\tfrac13 \theta^2+\frac{\left(c_a+1\right)
  \udot^2}{(3 c_{\theta }+1)}-\frac{2 \left(c_a+1\right) a \udot}{(3 c_{\theta }+1)}+\frac{6 \left(2
   c_{\sigma }-1\right)\sigma_+^2}{(3 c_{\theta }+1)}+\nonumber \\ &  +\frac{\hat{\mu} \left((\gamma -2) v^2-3 \gamma +2\right)}{2
   \left(3 c_{\theta }+1\right) \left(1-v^2\right)}, \\		
	&\e_0(\sigma_+)-\frac{\left(c_a+1\right) \e_1(\udot)}{3 \left(2 c_{\sigma }-1\right)}=-\theta\sigma_+ +\frac{1}{2}
   \sigma_+^2+\frac{\left(3 c_{\theta }+1\right) \theta^2}{18 \left(2 c_{\sigma }-1\right)}+\frac{\left(1-2
   c_a\right) a \udot}{3 \left(2 c_{\sigma }-1\right)}+\nonumber\\ & +\frac{\left(5 c_a+2\right) \udot^2}{6
   \left(2 c_{\sigma }-1\right)}-\frac{a^2}{2 \left(2 c_{\sigma }-1\right)}+\frac{K}{2 \left(2 c_{\sigma }-1\right)}+\frac{\hat{\mu}\left((\gamma +1) v^2-1\right)}{6 \left(2 c_{\sigma }-1\right) \left(1-v^2\right)},
\\
 &\e_0(a)+\frac{\left(3
   c_{\theta }+2 c_{\sigma }\right) \e_1(\theta)}{3(2 c_{\sigma }-1)}= -\tfrac13 (a+\udot)(\theta+3\sigma_+)-\frac{\gamma  \hat{\mu} v}{2 \left(2
   c_{\sigma }-1\right) (1-v^2)},\\
 &   \e_0(\hat{\mu}) -\frac{\e_1
  \left( \hat{\mu}   \right) v   \left(
 2-\gamma\right) }{G_-}-
 \frac {\gamma \hat{\mu}  \e_1 \left( v \right) }{G_-}  = -\frac {2\gamma
 \hat{\mu} v ^{2}
 \sigma_+  }{G_-} +\frac {
 \gamma  \left(v^{2}  -3\right)
 \hat{\mu}  \theta }{3 G_-}-2 \frac{\gamma \hat{\mu}  v a  }{G_-},
 \\
 \label{dv}
  &   \e_0(v) -{\frac {{
 \e_1} \left( \hat{\mu}   \right)  \left(1-v^2\right) ^{2}
  \left( \gamma-1 \right) }{ \gamma \hat{\mu} G_-  }}-
 {\frac {v   \left(2- \gamma\right) \e_1 \left(
 v   \right) }{G_-}} = {\frac {2 v   \left(1-v^2
  \right)  \sigma_+ }{G_-}}+ \nonumber \\ &{\frac {v
  \left(1-v^2\right)  \left( 3 \gamma-4 \right) \theta  }{ 3 G_-}}   +
 \frac {2 v ^{2} \left(1-v^2\right)  \left(
 \gamma-1 \right) a  }{G_-}+(1-v^2)\udot.
\end{align}
\end{subequations}
Constraints:
\begin{subequations}
\begin{align}
&  \e_1(\ln N) = \udot, \label{udot2}
\\ 
& \e_1(\ln K) = 2a,
\\
\label{eliminate_spatial2}
& \e_1(a)+c_a \e_1(\udot)=\frac{G_+ \hat{\mu}}{2    (1-v^2)}-\frac{1}{6}\left(3
   c_{\theta }+1\right) \theta^2-\frac{3}{2}\left(2 c_{\sigma }-1\right) \sigma_{+}^2+\nonumber \\ & -\frac{K}{2}+\frac{3 a^2}{2}+2 c_a a\udot-\frac{c_a \udot^2}{2},
\\\label{constraint_sigma_H}
& \e_1(\sigma_+)-\frac{\left(3 c_{\theta }+1\right) \e_1(\theta)}{3(2 c_{\sigma }-1)}= 3 a \sigma_+ +\frac{\gamma  \hat{\mu} v}{2 \left(2 c_{\sigma
   }-1\right) \left(1-v^2\right)}.
\end{align}
\end{subequations}


\subsection{Well-posedness}

We now show that the system of evolution equations plus restrictions for the state vector $$\left[\ex, K, \udot, \theta, \sigma_+, a, \hat{\mu}, v\right]^T$$ is well-posed for $\gamma \geq 1$. The coefficient matrix for the spatial derivative terms (for $\left[\udot, \theta, \sigma_+, a, \hat{\mu}, v\right]^T$) is:
\footnote{Strictly speaking, we should also include the factor $\ex$ in the
matrix, but the result on well-posedness is the same.}
\be
\label{matrix}
\left(
\begin{array}{cccccc}
 0 & -\frac{(3 c_\theta+2 c_\sigma)}{3c_a(2 c_\sigma-1)} & 0 & 0 & 0 & 0
   \\
 -\frac{c_a+1}{(3 c_\theta+1)} & 0 & 0 & 0 & 0 & 0 \\
 -\frac{c_a+1}{3(2 c_\sigma-1)} & 0 & 0 & 0 & 0 & 0 \\
 0 & \frac{3 c_\theta+2 c_\sigma}{3(2 c_\sigma -1)} & 0 & 0 & 0 & 0 \\
 0 & 0 & 0 & 0 & \dfrac{v(2-\gamma)}{(\gamma-1)v^2-1} & \dfrac{\gamma \hat{\mu}}{(\gamma-1)v^2-1} \\
 0 & 0 & 0 & 0 & \dfrac{(1-v^2)^2(\gamma-1)}{\left((\gamma-1)v^2-1\right)\gamma \hat{\mu}}& \dfrac{v(2-\gamma)}{(\gamma-1)v^2-1}
\end{array}
\right).
    \ee
		
Its eigenvalues are
\be
 0,0,\pm \sqrt{\frac{\left(c_a+1\right) \left(3 c_{\theta }+2 c_{\sigma }\right)}{3 c_a
   \left(3 c_{\theta }+1\right) \left(2 c_{\sigma }-1\right)}}, \; -\frac{(2-\gamma)v \pm \sqrt{\gamma-1}(1-v^2)}{G_-},
\ee
with corresponding eigenvectors (for example)
\begin{align}
& \left(\begin{array}{c}
 -\frac{1}{c_a} \\
 -\sqrt{\frac{\left(c_a+1\right) \left(6 c_{\sigma }-3\right)}{c_a \left(3 c_{\theta }+1\right) \left(3 c_{\theta }+2 c_{\sigma }\right)}} \\
 -\sqrt{\frac{\left(c_a+1\right) \left(3 c_{\theta }+1\right)}{c_a \left(3 c_{\theta }+2 c_{\sigma }\right) \left(6 c_{\sigma }-3\right)}} \\
 1 \\
 0 \\
 0 \\
\end{array}\right), \quad
\left(
\begin{array}{c}
 -\frac{1}{c_a} \\
 \sqrt{\frac{\left(c_a+1\right) \left(6 c_{\sigma }-3\right)}{c_a \left(3 c_{\theta }+1\right) \left(3 c_{\theta }+2 c_{\sigma }\right)}} \\
 \sqrt{\frac{\left(c_a+1\right) \left(3 c_{\theta }+1\right)}{c_a \left(3 c_{\theta }+2 c_{\sigma }\right) \left(6 c_{\sigma }-3\right)}} \\
 1 \\
 0 \\
 0 \\
\end{array}
\right),  \quad
	\left(\begin{array}{c}
 0 \\
 0 \\
 1 \\
 0 \\
 0 \\
 0 \\
\end{array}
\right),
	   \nonumber\\
	&	
    \left(\begin{array}{c}
 0 \\
 0 \\
 0 \\
 1 \\
 0 \\
 0 \\
\end{array}
   \right),\quad
    \left(
\begin{array}{c}
 0 \\
 0 \\
 0 \\
 0 \\
 \frac{\gamma  \hat{\mu} }{\left(1-v^2\right) \sqrt{\gamma -1}} \\
 1 \\
\end{array}
    \right)
		,\quad
    \left(
\begin{array}{c}
 0 \\
 0 \\
 0 \\
 0 \\
 -\frac{\gamma  \hat{\mu} }{\left(1-v^2\right) \sqrt{\gamma -1}} \\
 1 \\
\end{array}
    \right).
\end{align}

The matrix is diagonalizable (all its eigenvalues are real) for $\gamma > 1$ and $\frac{\left(c_a+1\right) \left(3 c_{\theta }+2 c_{\sigma }\right)}{c_a \left(3 c_{\theta }+1\right) \left(6 c_{\sigma }-3\right)}\geq 0$, with
$c_s = \sqrt{\gamma-1}$ being the speed of sound in the perfect fluid.
The system \eqref{synch}-\eqref{dv} is thus well-posed if  the above conditions are fulfilled, otherwise the system is elliptic and not well-posed.


\subsection{Normalized variables}

We introduce  the normalized variables (for $\udot\neq 0$ using the $\beta$-normalization):

$$\left\{\mathcal{N}^{-1}, E_{1}^{1},\mathcal{Q}, \Sigma, \mathcal{A}, \Udot \right\} = \left\{ N^{-1}, e_1^{1},\tfrac{\theta}{3}, \sigma_+ , a, \udot  \right\} / \beta $$
$$ \left \{\Omega, \Omega_k,  \mathcal{K},\calS \right \} = \left \{\hat{\mu}, -\tfrac{1}{2} \R, K, \S \right \} /(3 \beta^2),$$
where $\beta = \tfrac{1}{3}(\theta + 3\sigma_+)$ and 
$$ \R=4\e_1 a -6 a^2+2 K,\; \S=-\tfrac{1}{3}\e_1 a+\tfrac{1}{3}K.
$$
By definition $\mathcal{Q}+\Sigma=1.$ In the above we assume that  $\beta \neq 0$. In general the variables are
unbounded. However, 
physically $\Omega \geq 0$, and if the expansion and the shear are both positive, then $0 \leq \mathcal{Q} \leq 1$.

We also introduce the normalized differential operators,
$$ \pmb{\partial}_\alpha := \frac{\e_\alpha}{\beta} \quad;\  \text{where} \quad \alpha = 0, 1.$$
Moreover, we define $q$ and $r$ analogous to the usual volume deceleration parameter and ``Hubble spatial gradient" as follows
\begin{equation}
\pmb{\partial}_0 \beta := -(1+q)\beta, \quad \pmb{\partial}_1 \beta := -r\beta. 
\end{equation}  
so that
\begin{subequations}
\begin{align}
&\pmb{\partial}_0 \Ex\equiv \Ex (1+q)+\frac{1}{\beta^2} \e_0(\ex),~~
\pmb{\partial}_0 \mathcal{Q}\equiv \mathcal{Q} (1+q)+\frac{1}{3\beta^2} \e_0(\theta),\nonumber \\
&\pmb{\partial}_0 \Sigma\equiv \Sigma (1+q)+\frac{1}{\beta^2} \e_0(\sp),~\pmb{\partial}_0 \Udot\equiv \Udot (1+q)+\frac{1}{\beta^2} \e_0(\udot),~\pmb{\partial}_0 \mathcal{A}\equiv \mathcal{A} (1+q)+\frac{1}{\beta^2} \e_0(a),\nonumber\\
&\pmb{\partial}_0 \mathcal{K}\equiv 2\mathcal{K}(1+q)+\frac{1}{3\beta^3}\e_0(K),~
\pmb{\partial}_0 \Omega\equiv 2\Omega(1+q)+\frac{1}{3\beta^3}\e_0(\hat{\mu}),~
\pmb{\partial}_0 v\equiv\frac{1}{\beta} \e_0 v,\nonumber
\end{align}
\end{subequations} 
and
\begin{subequations}
\begin{align}
&\pmb{\partial}_1 \mathcal{Q}\equiv r \mathcal{Q} +\frac{1}{3\beta^2}\e_1(\theta),~
\pmb{\partial}_1 \Sigma\equiv r \Sigma +\frac{1}{\beta^2}\e_1(\sp),~
\pmb{\partial}_1 \Udot\equiv r \Udot +\frac{1}{\beta^2} \e_1(\udot),~
\pmb{\partial}_1 \mathcal{A}\equiv r \mathcal{A} +\frac{1}{\beta^2} \e_1(a),\nonumber\\
&\pmb{\partial}_1 \mathcal{K}\equiv  2 r \mathcal{K}+\frac{1}{3\beta^3}\e_1(K),~
\pmb{\partial}_1 \Omega\equiv 2 r \Omega+\frac{1}{3\beta^3}\e_1(\hat{\mu}),~\pmb{\partial}_1 v\equiv\frac{1}{\beta} \e_1 v.\nonumber
\end{align}
\end{subequations} 
Thus, the field equations reduce to 
\begin{subequations}
\label{(43)}
\begin{align}
& \pmb{\partial}_0 E_1^1  = (q + 3\Sigma) E_{1}^1,\label{eq96}\\
&  \pmb{\partial}_0 \mathcal{K}  = 2q \mathcal{K},\label{eq97}\\
& c_a\pmb{\partial}_0	\Udot-\frac{\left(3 c_{\theta }+2 c_{\sigma }\right) \pmb{\partial}_1 \mathcal{Q}}{ \left(2 c_{\sigma }-1\right)}=c_a\Udot(q-1)-\frac{r \left(3 c_{\theta }+2 c_{\sigma
   }\right) \mathcal{Q}}{\left(2c_{\sigma }-1\right)}+\frac{3 \gamma  c_{\sigma } v \Omega}{\left(2 c_{\sigma }-1\right)
   \left(1-v^2\right)},\label{eq98}\\
&  \pmb{\partial}_0 \mathcal{Q}-\frac{\left(c_a+1\right) \pmb{\partial}_1\Udot}{3 \left(3 c_{\theta
   }+1\right)}=\mathcal{Q} (1+q-\mathcal{Q})-\frac{r \left(c_a+1\right) \Udot}{3 \left(3 c_{\theta }+1\right)}-\frac{2 \left(c_a+1\right) \mathcal{A} \Udot}{3 \left(3 c_{\theta }+1\right)}+\nonumber \\
	& +\frac{\left(c_a+1\right) \Udot^2}{3 \left(3 c_{\theta
   }+1\right)}+\frac{2 \left(2 c_{\sigma }-1\right) \Sigma^2}{3 c_{\theta }+1}+\frac{\Omega \left((\gamma -2) v^2-3 \gamma
   +2\right)}{2 \left(3 c_{\theta }+1\right) \left(1-v^2\right)},\label{eq99}\\
&  \pmb{\partial}_0 \Sigma -\frac{\left(c_a+1\right)  \pmb{\partial}_1\Udot}{3(2 c_{\sigma
   }-1)} =\Sigma (1+q-3 \mathcal{Q})-\frac{r \left(c_a+1\right) \Udot}{3(2 c_{\sigma }-1)}+\frac{\left(1-2 c_a\right) \mathcal{A} \Udot}{3(2 c_{\sigma }-1)}+	\nonumber \\ & +\frac{3 \mathcal{K}}{2(2 c_{\sigma }-1)}+\frac{\left(5 c_a+2\right) \Udot^2}{6(2 c_{\sigma }-1)}-\frac{\mathcal{A}^2}{2(2 c_{\sigma
   }-1)}+\frac{\left(3 c_{\theta }+1\right) \mathcal{Q}^2}{2(2 c_{\sigma }-1)}+\frac{\Omega \left((\gamma +1) v^2-1\right)}{2 \left(2 c_{\sigma
   }-1\right) \left(1-v^2\right)}+\frac{1}{2} \Sigma^2,\label{eq100}\\	
& \pmb{\partial}_0 \mathcal{A}+\frac{\left(3 c_{\theta }+2 c_{\sigma }\right)  \pmb{\partial}_1 \mathcal{Q}}{2 c_{\sigma }-1}=q \mathcal{A}+\frac{r \left(3 c_{\theta }+2 c_{\sigma }\right)
   \mathcal{Q}}{2 c_{\sigma }-1}-\frac{3 \gamma  v \Omega}{2 \left(2 c_{\sigma }-1\right) \left(1-v^2\right)}-\Udot,\label{eq101}\\
& \pmb{\partial}_0 \Omega+\frac{(\gamma -2) v\pmb{\partial}_1\Omega}{G_-}-\frac{\gamma  \Omega \pmb{\partial}_1 v}{G_-}= -\frac{2 \gamma  \mathcal{A} v \Omega}{G_-} +\frac{2 q \Omega \left(1-(\gamma -1) v^2\right)}{G_-}+\nonumber \\ 
&-\frac{3 \gamma  \mathcal{Q}
   \left(1-v^2\right) \Omega}{G_-}+\frac{2 (\gamma -2) r v \Omega}{G_-}+\frac{2 \Omega
   \left((1-2 \gamma ) v^2+1\right)}{G_-},\label{eq102}	
\\
&\pmb{\partial}_0 v-\frac{(\gamma -1) \left(1-v^2\right)^2 \pmb{\partial}_1 \Omega}{\gamma 
  \Omega G_- }+\frac{(\gamma -2) v \pmb{\partial}_1 v}{G_-}= \frac{2 (\gamma -1) \mathcal{A} \left(1-v^2\right) v^2}{G_-}+\nonumber \\ & -\frac{3 (\gamma -2) \mathcal{Q} \left(1-v^2\right) v}{G_-}-\frac{2
   (\gamma -1) r \left(1-v^2\right)^2}{\gamma  G_-}+\frac{2 \left(1-v^2\right) v}{G_-}+\Udot \left(1-v^2\right), \label{eq103}
\end{align}
\end{subequations}
and
\begin{subequations}
\label{(44)}
\begin{align}
& \pmb{\partial}_1 \mathcal{N}^{-1} = (r - \dot{U}) \mathcal{N}^{-1}, \label{eq104} \\
& \pmb{\partial}_1 \mathcal{K} = 2(r + \mathcal{A})  \mathcal{K}, \label{eq105} \\
& \pmb{\partial}_1 \mathcal{A}+ c_a \pmb{\partial}_1 \Udot= -\frac{3}{2} \mathcal{K}+r \left(c_a \Udot+\mathcal{A}\right)+2 c_a \mathcal{A}\Udot-\frac{1}{2} c_a
   \Udot^2+\frac{3}{2} \mathcal{A}^2+\nonumber \\ & -\frac{3}{2}\left(3 c_{\theta }+2 c_{\sigma }\right) \mathcal{Q}^2
   +3\left(2 c_{\sigma }-1\right) \mathcal{Q}+\frac{3 \Omega \left((\gamma -1) v^2+1\right)}{2 \left(1-v^2\right)}-3 c_{\sigma }+\frac{3}{2}, \label{eq106a}\\
& \pmb{\partial}_1 \Sigma-\frac{\left(3 c_{\theta }+1\right)  \pmb{\partial}_1 \mathcal{Q}}{2 c_{\sigma }-1}=r \Sigma 
+3 \mathcal{A} \Sigma-\frac{r \left(3 c_{\theta }+1\right) \mathcal{Q}}{2 c_{\sigma }-1}+
\frac{3 \gamma  v \Omega}{2 \left(2 c_{\sigma }-1\right) \left(1-v^2\right)},\label{eq106}\\
&  \pmb{\partial}_0 \mathcal{A}  = (q + 3\Sigma)\mathcal{A} - \Udot + r.\label{eq107}
\end{align}
\end{subequations}

Note that $\mathcal{A}^2, \mathcal{K}$, only appear in the equations \eqref{eq99}, \eqref{eq100}, \eqref{eq101} and  \eqref{eq106a} via the combination  
$\mathcal{D} \equiv \mathcal{A}^2 -3\mathcal{K}$ (and in the other equations also via terms of the form  
$\mathcal{A}\dot{U}, \mathcal{A}v$, the definition of $q$ \eqref{def_q}, etc.).
We further develop the governing equations; however, since this is rather technical
we continue this development in Appendix B.


\section{Dust models}

Let us consider dust models with $\gamma=1$ ($p=0$).
The special case of dust, in which the governing equations simplify considerably, is
of particular interest at late times.
In GR we immediately obtain the simple Lema\^{\i}tre-Tolman-Bondi (LTB) model with $\dot{u}=0$
(and since $N=N(t)$ is a function of $t$, we can set $N=1$ by a time rescaling), 
where the fluid is ``comoving'' ($v = 0$); i.e.,  $\dot{u}=0$ and $v = 0$
simultaneously (see Appendix C). This is not possible in the models here; in general $v$ cannot be zero 
in the dust case and
there is no GR-like ``LTB'' model.

If $v = 0$ ($\gamma=1$), from equations \eqref{dudot} -\eqref{dv}  we immediately find that  $\dot{u}=0$,
which is a contradiction (for equations \eqref{dudot} -\eqref{dv}). Therefore, in our formalism, the perfect fluid must be
tilting ($v\neq 0$). In general, we thus need to investigate dust with $v\neq 0$ and $\dot{u} \neq0$ (i.e., non-``LTB''). 
Let us next consider the case   $\dot{u}=0$ (see equations (\ref{udoteqns1})-(\ref{udoteqns1b})). From equations 
(\ref{udoteqns2})-(\ref{udoteqns2b}), if $v = 0$ 
we then find that (either) $3c_\theta+2c_\sigma=0$ ($c_\theta=c_\sigma=0$ in GR)
(or $\e_1(\theta)=0$, which is valid in the spatially homogeneous models); we could investigate this special 
model further.

Let us also consider the subcase  $\dot{U}=0$ and $v = 0$ in normalized variables.

\subsection{Normalized equations}

Let us study the special subset  $\dot{U}=v=0$, with $\gamma = 1$ (see Appendix B).
We also assume that  $c_\theta=0$ and $c_\sigma \neq 0$: 
 \begin{subequations}
\begin{align}
&  \pmb{\partial}_0 \mathcal{K}  = 2q \mathcal{K}, \\
&  \pmb{\partial}_0 \mathcal{Q}=\mathcal{Q} (1+q-\mathcal{Q})+{2 \left(2 c_{\sigma }-1\right) (1-\mathcal{Q})^2}-
\frac{1}{2} \Omega,\\
& \pmb{\partial}_0 \mathcal{A}=q \mathcal{A},\\
& \pmb{\partial}_0 \Omega= (2 q -3 \mathcal{Q} +2)\Omega,
\end{align}
\end{subequations}
subject to the restrictions: 
\begin{subequations}
\begin{align}
& \pmb{\partial}_1 \mathcal{K} = 2(r + \mathcal{A})  \mathcal{K}, \\
& \pmb{\partial}_1 \mathcal{A}= -\frac{3}{2} \mathcal{K}+r \mathcal{A}+\frac{3}{2} \mathcal{A}^2 -3 c_{\sigma }\mathcal{Q}^2
+\frac{3}{2}\left(2 c_{\sigma }-1\right)(2 \mathcal{Q} -1) +\frac{3 }{2}\Omega,\\
& \pmb{\partial}_1 \mathcal{Q}=r \mathcal{Q}, 
\end{align}
\end{subequations}
where $q$ and $r$ are defined by:

\begin{align}
& q= \frac{1}{2(2c_{\sigma }-1)}\Big\{-3\mathcal{K}+\mathcal{A}^2-2 c_{\sigma }\left(8c_{\sigma }-3\right) \mathcal{Q}^2 
+ 16c_{\sigma }(2c_{\sigma }-1)\mathcal{Q} \nonumber \\ 
& +2c_{\sigma }\Omega +(1-8 c_{\sigma })(2c_{\sigma }-1)\Big\}.
\end{align}
\begin{align}
r=- 3 \mathcal{A} (1- \mathcal{Q}).
\end{align}
We also have that
$ \pmb{\partial}_0 :=  \mathcal{N}^{-1} \partial_t$,\quad
$ \pmb{\partial}_1 := E_{1}^1 \partial_x$.
[The only remaining freedom is the coordinate rescalings $t \rightarrow f(t)$ and 
$x \rightarrow g(x)$].

\subsection{Special dust model}

The terms $\mathcal{K}, \mathcal{A}^2$, only appear in the evolution equations for $\pmb{\partial}_0\Omega, \pmb{\partial}_0\mathcal{Q}$
via (through $q$) the combination $\mathcal{D} \equiv \mathcal{A}^2 -3\mathcal{K}$. Hence (assuming
$c_{\sigma } \neq 0$) we have the reduced (closed) evolution system:

\begin{subequations}
	\label{syt70}
\begin{align}
&  \pmb{\partial}_0 \mathcal{D}  = 2q \mathcal{D}, \\
&  \pmb{\partial}_0 \mathcal{Q}=\mathcal{Q} (1+q-\mathcal{Q})+{2 \left(2 c_{\sigma }-1\right) (1-\mathcal{Q})^2}-
\frac{1}{2} \Omega,\\
& \pmb{\partial}_0 \Omega= (2 q -3 \mathcal{Q} +2)\Omega,
\end{align}
\end{subequations}
where
\begin{align}
& q= \frac{1}{2(2c_{\sigma }-1)}\Big\{\mathcal{D}-2 c_{\sigma }\left(8c_{\sigma }-3\right) \mathcal{Q}^2 
+ 16c_{\sigma }(2c_{\sigma }-1)\mathcal{Q} \nonumber \\ 
& +2c_{\sigma }\Omega +(1-8 c_{\sigma })(2c_{\sigma }-1)\Big\}.
\end{align}

In the decoupled evolution equations above 
(which are only valid strictly speaking for  $c_\sigma \neq 0$)
we have not yet applied any constraints.
The constraint eqns for ``LTB''-like models (i.e., dust models in Einstein-aether theory with $\dot{U}=0$ and $v = 0$)
imply either $3c_\theta+2c_\sigma=0$ or $\e_1(\theta)=0$.
The problems regarding ``LTB'' come from the constraints for $\e_1(\theta)$ and $\e_1(\sigma_{+})$; when normalizing with $3\beta=
\theta+3\sigma_{+}$, these constraints get hidden in the normalized variables (because $\theta$ decouples in the normalized eqns and $\Sigma$
is related to $\sigma_{+}/(\theta+3\sigma_{+}))$, and so there are no problems per se with normalized equations But they do not represent
any ``LTB'' model because the constraints are not satisfied.

The FLRW models in this special dust model (with
$c_\sigma \neq 0$) have $\mathcal{D}=0, \Omega=1, 
\mathcal{Q}=1$ ($q = \frac{1}{2}$) and correspond to an equilibrium point (the point $P_4$ below).
In the Kantowski-Sachs models, $\mathcal{A}=0, r=0$, and there are no spatial derivatives,
and the constraints can be used to eliminate the (non-zero) $\mathcal{D}$ and the
resulting system becomes 2-dimensional (the Kantowski-Sachs models will be studied later using a different
normalization).

Assuming $\mathcal{A}\neq 0$, we can define the new spatial derivative 
$\pmb{\partial}_\eta\equiv \mathcal{A}^{-1} \pmb{\partial}_1$, whence the spatial derivatives become:
		\begin{subequations}
			\begin{align}
			&\pmb{\partial}_\eta \mathcal{D}= 3\left(-2 c_\sigma (1-\mathcal{Q})^2+(\mathcal{D}-1)(-1+2 \mathcal{Q})+\Omega\right),\\
			&\pmb{\partial}_\eta \mathcal{Q}= -3(1-\mathcal{Q})\mathcal{Q}.
			\end{align}
		\end{subequations}
				The commutator equation is given by 
		\begin{equation}
		\left[\pmb{\partial}_\eta, \pmb{\partial}_\tau\right]=\mathcal{A}^{-1}\left(q \pmb{\partial}_0+\left[\pmb{\partial}_1, \pmb{\partial}_0\right]\right).
		\end{equation}
		
There is no spatial restriction for $\Omega$; thus, it is freely specified at the initial spatial hypersurface.


Let us summarize the equilibrium points of the system (\ref{syt70}) and their eigenvalues (see Table \ref{Tab70}), 
and discuss their stability:
 
 		\begin{table}
 			\renewcommand{\arraystretch}{1.95}
 			{%
 				\begin{tabular}{|l|c|c|}
 					\hline
 					Label & $(\mathcal{D},\mathcal{Q},\Omega)$	& Eigenvalues \\	\hline
 					$P_{1}$ & $\left(0, 1+\frac{\sqrt{1-2 c_{\sigma }}}{2 c_{\sigma }}-\frac{1}{2 c_{\sigma }}, 0\right)$	& $\frac{4 c_{\sigma }+3 \sqrt{1-2 c_{\sigma }}-3}{c_{\sigma }},\frac{8 c_{\sigma }+3 \sqrt{1-2 c_{\sigma }}-3}{2 c_{\sigma }},\frac{3 \left(2 c_{\sigma
 						}+\sqrt{1-2 c_{\sigma }}-1\right)}{2 c_{\sigma }}$  \\[1mm]
 					$P_{2}$ & $\left(0, 1-\frac{\sqrt{1-2 c_{\sigma }}}{2 c_{\sigma }}-\frac{1}{2 c_{\sigma }}, 0\right)$ & $\frac{4 c_{\sigma }-3 \sqrt{1-2 c_{\sigma }}-3}{c_{\sigma }},-\frac{-8 c_{\sigma }+3 \sqrt{1-2 c_{\sigma }}+3}{2 c_{\sigma }},-\frac{3 \left(-2
 						c_{\sigma }+\sqrt{1-2 c_{\sigma }}+1\right)}{2 c_{\sigma }}$ \\[1mm]
 					$P_{3}$ & $\left(0, \frac{4 \left(2 c_{\sigma }-1\right)}{8 c_{\sigma }-3}, 0\right)$ & $-\frac{3}{8 c_{\sigma }-3},-\frac{32 c_{\sigma }-15}{2 \left(8 c_{\sigma }-3\right)},-1$\\[2mm]
 					$P_{4}$ & $(0,1,1)$ & $1, 1, -\frac{3}{2}$\\[2mm]
 					$P_{5}$ & $(1-2c_{\sigma},\frac{2}{3},\frac{8c_{\sigma}}{9})$ & $1,\frac{-3+\sqrt{9-48c_{\sigma}}}{6},\frac{-3-\sqrt{9-48c_{\sigma}}}{6}$ \\[1mm]
 						$P_{6}$ & $(1,1,0)$ & $-1,-1-\sqrt{\frac{2 c_{\sigma }}{2 c_{\sigma }-1}},-1+ \sqrt{\frac{2 c_{\sigma }}{2 c_{\sigma }-1}}$\\[2mm]
 						$P_{7}$ & $\left(\frac{3 \left(2 c_{\sigma }-1\right) \left(8 c_{\sigma }-3\right)}{\left(4 c_{\sigma }-3\right){}^2},  \frac{2 \left(2 c_{\sigma
 							}-1\right)}{4 c_{\sigma }-3}, 0\right)$ & $-\frac{4c_{\sigma}}{4c_{\sigma}-3},-\frac{3}{4c_{\sigma}-3},-\frac{8c_{\sigma}-3}{4c_{\sigma}-3}$\\[2mm]
 				 					\hline	
 				\end{tabular}
 			}
 			\caption{\label{Tab70} Equilibrium  points and eigenvalues of the system \eqref{syt70}.}
 		\end{table}

 \begin{enumerate}
 	\item Point $P_1$ exists (i.e., with $-1\leq Q\leq 1$) for $c_{\sigma }<0$ 
 	or $0<c_{\sigma }\leq \frac{1}{2}$. It is a source for $c_{\sigma }<0$ or $0<c_{\sigma }<\frac{3}{8}$; 
 	a saddle for $\frac{3}{8}<c_{\sigma }<\frac{1}{2}$. (The equilibrium points are non-hyperbolic for
 	other values of the parameter $c_{\sigma }$). 
 	\item Point $P_2$ exists for $\frac{3}{8}\leq c_{\sigma }\leq \frac{1}{2}$. 
 	It is a sink for $\frac{3}{8}\leq c_{\sigma }<\frac{15}{32}$;  a saddle for 
 	$\frac{15}{32}<c_{\sigma }<\frac{1}{2}$. 
 	\item Point $P_3$ exists for $c_{\sigma }\geq \frac{7}{16}$. 
 	It is a sink for $c_{\sigma }>\frac{15}{32}$; a saddle for $\frac{7}{16}\leq c_{\sigma }<\frac{15}{32}$. 
 	\item The [FLRW] point $P_4$ always exists and it is a saddle. 
 	\item The point $P_5$ always exists and it is a saddle for $c_{\sigma }\neq 0$. 
 	\item The point $P_6$ always exists. It is sink for $c_{\sigma }<\frac{1}{2}$ 
 	[two complex conjugate eigenvalues with negative real part for $0<c_{\sigma }<\frac{1}{2}$]. 
 	It is a saddle for $c_\sigma>\frac{1}{2}$. For $c_\sigma=\frac{1}{2}$ it is a saddle too.
 	\item The point $P_7$ exists for $c_{\sigma }\leq \frac{5}{8}.$ It is a source for $\frac{3}{8}<c_{\sigma }\leq \frac{5}{8}$. 
 	Non-hyperbolic for $c_\sigma\in\{0, \frac{3}{8}\}$. Saddle otherwise. 
 \end{enumerate}

 \begin{figure}
 	\centering
 	\includegraphics[width=0.6\linewidth]{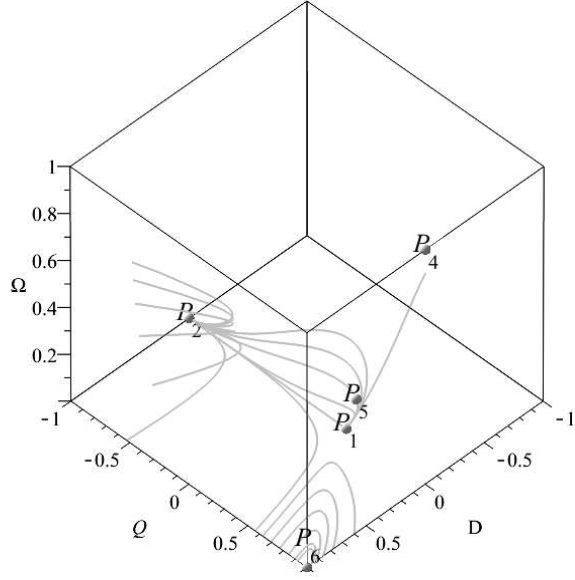}
 	\caption{Phase space of the system \eqref{syt70} for the choice $c_\sigma=\frac{3}{8}$. 
 	The sinks are $P_2$ and $P_6$. $P_7$ and $P_1$ coincide; they are non-hyperbolic and behave as the sources.}
 	\label{fig:Syst69_c_0.375}
 \end{figure}

 \begin{figure}
 	\centering
 	\includegraphics[width=0.6\linewidth]{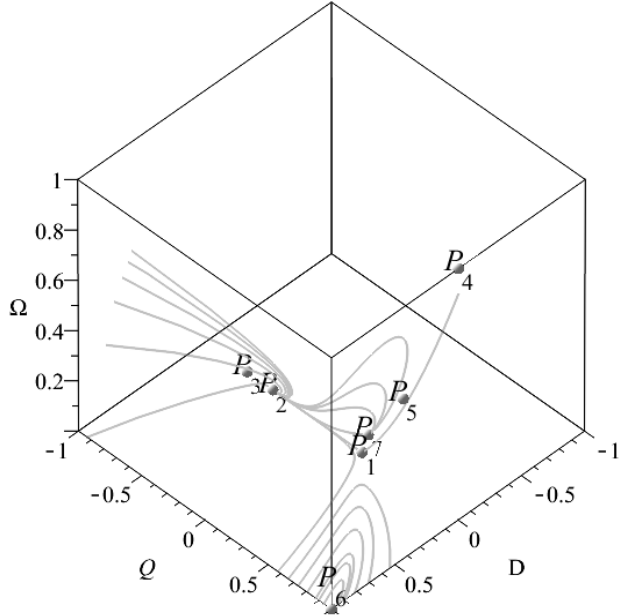}
 	\caption{Phase space of the system \eqref{syt70} for the choice 
 	$c_\sigma=0.45\in \left(\frac{3}{8}, \frac{15}{32}\right)$. The sinks are $P_2$ and $P_6$, and  $P_7$ is the source.}
 	\label{fig:Syst69_c_0.45}
 \end{figure}

 \begin{figure}
 	\centering
 	\includegraphics[width=0.6\linewidth]{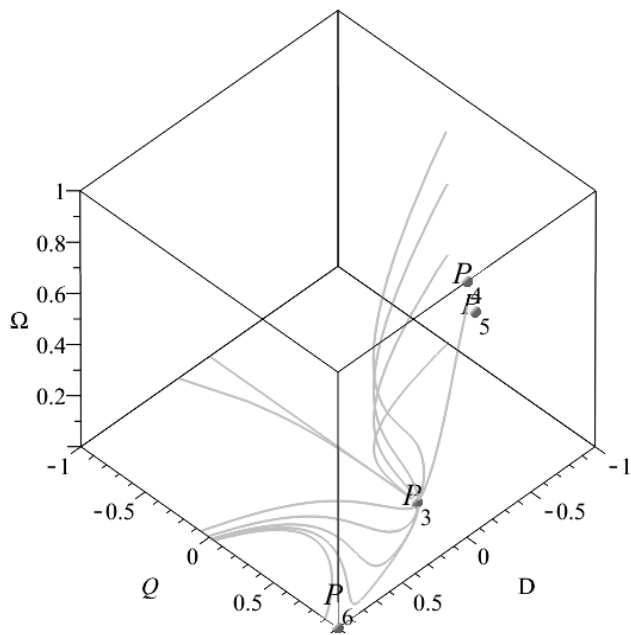}
 	\caption{Phase space of the system \eqref{syt70} for the choice $c_\sigma=0.7$. The sink is $P_3$.}
 	\label{fig:Syst69_c_0.7}
 \end{figure}
 

{\it{Discussion:}} Let us define 
\begin{align}
{D}_{\sigma} =  (1-8 c_{\sigma })(1-2c_{\sigma })
+ 16c_{\sigma }(1-2c_{\sigma })\mathcal{Q} 
-2 c_{\sigma }\left(3-8c_{\sigma }\right)\mathcal{Q}^2, \nonumber 
\end{align}
and consider the equilibrium points at finite values:

\noindent (a) $\mathcal{D}=0, \Omega=0,
\mathcal{Q}= Q_{a}$ (where there are constraints on the parameter 
$c_{\sigma }$ in order for  $Q_{a}$ to be physical)  
[{the points $P_1, P_2, P_3$ in Table \ref{Tab70}}]; generically a saddle.

\noindent (b) 
$\mathcal{D}=0, \Omega=1, 
\mathcal{Q}= 1$ ($q = \frac{1}{2}$). Eigenvalues:
$\lambda_1 = 1$, $\lambda_2=1$ and $\lambda_3=-\frac{3}{2}$   
The FLRW equilibium point [{point $P_4$ in Table \ref{Tab70}}] is always a saddle.

\noindent (c) 
$\mathcal{D}= 1-2 c_\sigma,
\Omega=\frac{8}{9}c_{\sigma }, 
\mathcal{Q}= \frac{2}{3}$ ($q = 0$) 
[{point $P_5$ in Table \ref{Tab70}.}]
 
\noindent (d) 
$\mathcal{D}={D}_{\sigma}, \Omega=0$
($q = \frac{1}{2}$), and  either (i)
$\mathcal{Q}= 1$ ($\mathcal{D}=1$; no shear) or (ii)
$\mathcal{Q}= Q_d =  2(1-2c_{\sigma })/(3-4c_{\sigma })$ 
($\mathcal{D}=3(1-2c_{\sigma })(3-8c_{\sigma })/(3-4c_{\sigma })^2$).
Eigenvalues: (i)
($\mathcal{Q}= 1$ [{point $P_6$ in Table \ref{Tab70}])
$\lambda_1 = -1$,   $\lambda_{2,3} = - 1 \pm \sqrt{\frac{-2c_{\sigma }}{(1-2c_{\sigma }) }}$ 
(negative real part for 
$c_{\sigma } < \frac{1}{2}$) -- corresponding to a sink, (ii) [{point $P_7$ in Table \ref{Tab70}}]
$\lambda_1 = \frac{4c_{\sigma }}{(3-4c_{\sigma })}$, $\lambda_2 = -1 + O(c_{\sigma })$,
$\lambda_3 = 1 + O({c_\sigma })$ -- which is a saddle for small ${c_\sigma }$.

{\em{Summary of sinks:} $P_6$ for $c_{\sigma }<\frac{1}{2}$, 
$P_2$  for $\frac{3}{8}\leq c_{\sigma }<\frac{15}{32}$, 
$P_3$  for $c_{\sigma }>\frac{15}{32}$. In all cases $\Omega \rightarrow 0$
to the future. For $P_2$ and $P_3$, $\mathcal{D} \rightarrow 0$, but for 
$P_6$, $\mathcal{D} \rightarrow 1$ ($\mathcal{Q} \rightarrow 1$) and the shear goes to zero at late times
(for small  $c_{\sigma }$). There is a range of values of the parameter $c_{\sigma }$ for which the 
sinks $P_2$ and $P_3$ represent inflationary solutions.
}


For illustration, we present some phase portraits
of the system \eqref{syt70} for the  $x$-constant surfaces in figures 
\ref{fig:Syst69_c_0.375} - \ref{fig:Syst69_c_0.7}.
In figure \ref{fig:Syst69_c_0.375} we present some orbits of the 
phase space for the parameter $c_\sigma=\frac{3}{8}$. The sinks are $P_2$ and $P_6$. $P_7$ and $P_1$ 
coincide; they are non-hyperbolic and behave as the sources. 
In figure \ref{fig:Syst69_c_0.45} we present the evolution of the system \eqref{syt70} 
for $c_\sigma=0.45\in \left(\frac{3}{8}, \frac{15}{32}\right)$. The sinks are $P_2$ and $P_6$.  $P_7$ is the source.
Finally, in figure \ref{fig:Syst69_c_0.7} we present 
the phase portrait for the choice $c_\sigma=0.7$. The sink is $P_3$.


\section{Special cases with extra Killing vectors}

Spherically symmetric models with more than 3 KVF are either
spatially homogeneous or static. 
Spatially homogeneous spherically symmetric models are
either Kantowski-Sachs models, or the
 Friedmann-Lema\^{\i}tre-Robertson-Walker (FLRW) models (with or without cosmological constant
 $\Lambda$)  [or locally rotationally symmetric (LRS) Bianchi I and Bianchi III models].
For the FLRW and Kantowski-Sachs models we use the equations in the case that $\dot{u}=0$, which follows 
immediately from the condition that $N=N(t)$.
 Static and self-similar spherically symmetric models have been studied in
\cite{static,SSSS}. \footnote{Recall that we have chosen a gauge so that the aether is aligned with $\e_0$.}

\subsection{The FLRW models}

 For the FLRW models the source must be of the form of a comoving perfect fluid (or vacuum)
 and the aether must be comoving.
  The metric has the form
 \be
         ds^2 = - N(t)^2 dt^2 + \ell^2(t) dx^2
                 + \ell^2(t) f^2(x) (d\y^2 + \sin^2 \y  d\z^2),
 \ee
 with
 \be
 \label{fx_FLRW}
     f(x) = \sin x,\ x,\ \sinh x,
 \ee
 for closed, flat, and open FLRW models, respectively.
 The frame coefficients are given by $\ex = \ell^{-1}(t)$ and $\ey =
 \ell^{-1}(t) f^{-1}(x)$. Then $\sp = \frac13\e_0 \ln(\ex/\ey)$
 vanishes. Furthermore, $a=-\frac{\partial_x f(x)}{f(x) \ell(t)}.$
 $N=N(t)$ implies that $\udot=0$; i.e.,
 the temporal gauge is synchronous, and we can set $N$ to any positive
 function of $t$ (we usually choose  $N=1$).
 The Hubble scalar $H = \e_0 \ln \ell(t)$ is also a function of
 $t$.
 {\footnote {We shall not list the KVFs as they are complicated in spherically
 symmetric coordinates and not needed here.}}

 For the spatial curvatures, $\S$ does vanish because (\ref{fx_FLRW})
 implies $\e_1 a = K$,%
 \footnote{That $\e_1 a$ does not vanish is consistent with
  the frame vector $\e_1$ not being group-invariant.}
  while $\R$ simplifies to
 \be
     \R = \frac{6\kappa}{\ell^2},\quad
     \kappa = 1,0,-1,
 \ee
 for closed, flat, and open FLRW, respectively.
 The evolution equation for $\sp$ and the Codazzi constraint then imply
 that $\pi_+ =0= q_1.$

FLRW cosmological models with aether and a comoving perfect fluid comoving have been studied 
previously   \cite{Jacobson,Jacobson:2000xp,Zlosnik:2006zu,Carroll:2004ai}. It was found that 
there is no essential affects on standard cosmology
in the minimal aether theory.
FLRW cosmological models  with a scalar field were studied in \cite{DJ,Barrow:2012qy,Sandin}.
The decay of  tilt has  
also been studied in (anisotropic and non-comoving) models  with $\Lambda$ \cite{CarrJ,kann}.

 
\subsubsection{The FLRW models in normalized coordinates}

The FLRW models in normalized coordinates are characterized by $\Sigma =0$ ($Q=1$), $\dot{U}=0$, $v=0$,
and we can use the remaining coordinate freedom to set $N=1$ (where $\beta =  \frac{\dot{\ell}}{N \ell}$).
We recall that
$ \pmb{\partial}_1 := E_{1}^1 \partial_x$ 
and $\pmb{\partial}_0 E_1^1  = q E_{1}^1$.
We use the remaining spatial  freedom
to simplify $f(x)$ as in equation (\ref{fx_FLRW})
for the FLRW metric as above, where $({\partial_x}{f(x)})^2 \equiv 1 - \kappa f^2$,   ${\partial_x}{\partial_x}{f(x)} \equiv  
- \kappa f$,  and so we obtain:~\footnote{Flat FLRW power law models:
In the flat case $\kappa=0$, $f(x)=x$ (and $\tau = \ln(\ell)$).
At the equilibrium points we have that $\ell = t^p$.}
\begin{align} \label{FLRWa}
& \mathcal{K} = \frac{N^2}{3 \dot{\ell}^2 f^2}
\end{align}
\begin{align}\label{FLRWb}
&  \mathcal{A} = -\frac{N{\partial_x}{f}}{\dot{\ell} f},
\end{align}
and hence
\begin{align}\label{FLRWc}
& \mathcal{A}^2  - 3 \mathcal{K} = -\frac{\kappa N^2}{{\dot{\ell}}^2},
& \pmb{\partial}_1 \mathcal{A} = \frac{3}{N}\mathcal{K}.
\end{align}

\subsubsection{The subset $\dot{U}=v=0$ with $Q=1$} 
We take the equations in the case $\dot{U}=v=0$ presented earlier, and set $Q=1$.
We again assume that  $c_\theta=0$ and $c_\sigma \neq 0$ (and, in principle, $\gamma \neq 1$). 
Since $Q=1$,  $\Sigma =0$ (i.e., the shear is zero), which is not in general an invariant set.
We immediately have that $r=0$, whence  $\pmb{\partial}_1 \mathcal{N}^{-1}=0$, and we can
 rescale time so that $N=1$ and $ \pmb{\partial}_0 := \partial_\tau$, where $\tau$ is essentially logarithmic time.
 We also have that $\Omega$ is independent of space, and that

\begin{subequations}
	\label{eq78}
\begin{align}
& \pmb{\partial}_\tau E_1^1  = q  E_{1}^1, \\
&  \pmb{\partial}_\tau \mathcal{K}  = 2q \mathcal{K}, \\
& \pmb{\partial}_\tau \mathcal{A}=q \mathcal{A},\\
& \pmb{\partial}_\tau \Omega= (2 q -3 \gamma  +2)\Omega,
\end{align}
\end{subequations}
subject to the restrictions: 
\begin{subequations}
	\label{eq79}
\begin{align}
& \pmb{\partial}_1 \mathcal{K} = 2 \mathcal{A}  \mathcal{K}, \\
& \pmb{\partial}_1 \mathcal{A}= -\frac{3}{2} \mathcal{K}+\frac{3}{2} \mathcal{A}^2 +\frac{3 }{2}(\Omega - 1),\\
\end{align}
\end{subequations}
where $q$ is defined by:
\begin{align}
\label{eq80}
&  q=\frac{1}{2} \Omega \left(3 \gamma - 2\right),
\end{align}
and where
\begin{align}
\label{eq81}
& 0=-3\mathcal{K}+\mathcal{A}^2 + \Omega  -1.
\end{align}

Note that if we differentiate this constraint and use the  constraint and the definition of $q$,
we obtain zero; hence the constraint is conserved along the evolution.
We can use this constraint to eliminate $\mathcal{K}$ from the above equations.
We note, as expected, that all dependence on $c_\sigma$ has dropped out. 
We also note that the above system has the equilibrium points $\Omega=0, q=0$, corresponding to late time vacuum, and 
$\Omega=1, q=\frac{1}{2}(3 \gamma - 2)$, the early time flat solution.
 
Finally, using equations (\ref{FLRWa}, \ref{FLRWb} \ref{FLRWc}), we obtain
\begin{equation}
\pmb{\partial}_\tau \Omega= (2 q -3 \gamma   +2)\Omega,
\end{equation}
where $q=\frac{1}{2} \Omega \left(3 \gamma - 2\right)$ and $ \Omega = 1  + \frac{\kappa}{{\dot{\ell}}^2}$,
as expected.

Note that in the general solution (i.e., not FLRW)
we can define $\mathcal{D}=\mathcal{A}^2-3\mathcal{K}$, so that
$\pmb{\partial}_\tau \mathcal{D}  = 2q \mathcal{D}$,
subject to the restrictions: $\pmb{\partial}_\eta \mathcal{D}= 3\left(\mathcal{D} + \Omega  -1\right)$,
where we have introduced the new spatial coordinate
$\pmb{\partial}_\eta \equiv \pmb{\partial}_1/\mathcal{A}$.
Since $0=\mathcal{D} + \Omega  -1$
we then obtain $\pmb{\partial}_\eta \mathcal{D}=0$, which implies $\pmb{\partial}_\eta \Omega=0$.
We then obtain 
 		\begin{align*}
 		 		&\mathcal{D}=\frac{1}{e^{-3 \gamma  \tau -c_1+2 \tau }+1}, ~\Omega=\frac{e^{2 \tau }}{e^{3 \gamma  \tau +c_1}+e^{2 \tau }}, 
 		 		~ q= \frac{(3 \gamma -2) e^{2 \tau }}{2 \left(e^{3 \gamma  \tau +c_1}+e^{2 \tau }\right)}, \\ & \mathcal{A}=\frac{c_2(\eta) e^{\frac{3 \gamma  \tau }{2}}}{\sqrt{e^{3 \gamma  \tau +c_1}+e^{2 \tau }}},
 		 		~\mathcal{K}=\frac{e^{3 \gamma  \tau } \left(c_2(\eta )^2-e^{c_1}\right)}{3 \left(e^{3 \gamma  \tau +c_1}+e^{2 \tau }\right)}.
 		\end{align*}
 	The equations 
 	\begin{subequations}
 		\begin{align}
 		&\pmb{\partial}_\eta \mathcal{K}  = 2 \mathcal{K},\\
 		&\mathcal{A} \pmb{\partial}_\eta \mathcal{A}=-\frac{3}{2} \mathcal{K}+\frac{3}{2} \mathcal{A}^2 +\frac{3 }{2}(\Omega - 1),
 		\end{align}
 	\end{subequations}
 	 	are identically satisfied if
 		\begin{align}
 		e^{c_1}+c_2(\eta ) \left(c_2'(\eta )-c_2(\eta )\right)=0, c_2(\eta )\neq 0.
 		\end{align}
 	The above equations admit the solutions
 		\begin{align*}
 		&c_2(\eta )=\pm\sqrt{e^{c_1}-e^{2 c_2+2 \eta }},\;
 		E_1^1 =\frac{c_3(\eta ) e^{\frac{3 \gamma  \tau }{2}}}{\sqrt{e^{3 \gamma  \tau +c_1}+e^{2 \tau }}},\\
 		&\mathcal{K}=-\frac{e^{3 \gamma  \tau +2 c_2+2 \eta }}{3 \left(e^{3 \gamma  
 		\tau +c_1}+e^{2 \tau }\right)},\; 
 		\mathcal{A}=\pm\frac{\sqrt{e^{c_1}-e^{2 \left(c_2+\eta \right)}}
 			e^{\frac{3 \gamma  \tau }{2}}}{\sqrt{e^{3 \gamma  \tau +c_1}+e^{2 \tau }}}.
 		\end{align*}

 
\subsection{The Kantowski-Sachs models}

We now investigate the spatially homogeneous subcase, in which a full global analysis
is possible. It is of particular interest whether general solutions can asymptote towards 
spatially homogeneous solutions
at late or early times. The spatially homogeneous spherically symmetric models (that has 4
 Killing vectors, the fourth being $\partial_x$) are the so-called
 Kantowski-Sachs models \cite{kramer}.
We shall consider the special comoving aether case.
 The metric (\ref{metric}) simplifies to
 \be
         ds^2 = - N(t)^2 dt^2 + (\ex(t))^{-2} dx^2
                 + (\ey(t))^{-2} (d\y^2 + \sin^2 \y  d\z^2);
 \ee
i.e., $N$, $\ex$ and $\ey$ are now independent of $x$.
The spatial derivative terms $\e_1(\ )$ vanish and as a
result $a=0=\udot$. Since $\udot=0$,  $N$ is a positive function of $t$ which under a time rescaling can be set to one.
This metric choice forces the fluid to be non-tilted ($v=0$) [assuming $\mu>0, \gamma>0$]. 

The evolution equations for the Kantowski-Sachs metric for an Einstein-aether spherically symmetric cosmology, in the presence of a perfect fluid, are: 
 \begin{subequations}
 \begin{align}
 &\e_0 (\ex) =-\tfrac13 (\theta-6\sigma_+) \ex
 \\
    & \e_0 (K) = -\tfrac23(\theta+3\sigma_+)K\\
 &\e_0 (\theta)=-\frac{\theta^2}{3}+\frac{6 (2 c_\sigma-1) \sigma_+^2}{3 c_\theta +1} +\frac{\left(2- 3 \gamma \right)\hat{\mu}}{2(3 c_\theta+1)},\\
 &\e_0(\sigma_+)= -\frac{(3 c_\theta +1)\theta^2}{9 (2c_\sigma-1)}-\theta\sigma_+-\sigma_+^2+\frac{\hat{\mu}}{3  (2c_\sigma-1)},\\
 &\e_0 (\hat{\mu})=-\gamma \theta \hat{\mu}
 \end{align}
\end{subequations}
 with the constraint
 \begin{equation}
 K+\frac{(3 c_\theta +1)\theta^2}{3}=\hat{\mu}-3  (2c_\sigma-1)\sigma_+^2.
 \end{equation}

We choose the following normalized variables (which are bounded for $1-2 c_\sigma\geq 0$; note that we do not use 
the $\beta-$normalization for convenience here):
 \begin{equation}\label{KS_vars_1}
 x=\frac{\sqrt{\hat\mu}}{D}, y=\frac{\sqrt{3}\sigma_+}{D}, z=\frac{\sqrt{K}}{D}, Q=\frac{\theta}{\sqrt{3}D}
 \end{equation}
 \noindent
 where\\
 \begin{equation}
 D=\sqrt{K+\frac{\theta^2}{3}},
 \end{equation}
and the new time variable $f'\equiv\frac{1}{D} \e_0 (f).$

We then obtain the full 4 dimensional (4D) system:
\begin{subequations}
\begin{align}
& x'=x \left(\frac{Q y^2 \left(2-4 c_{\sigma }\right)}{\sqrt{3} \left(3 c_{\theta }+1\right)}+\frac{Q^3}{\sqrt{3}}+\frac{Q \left(2 z^2-3 \gamma \right)}{2
   \sqrt{3}}+\frac{y z^2}{\sqrt{3}}\right)+\frac{(3 \gamma -2) Q x^3}{2 \sqrt{3} \left(3 c_{\theta }+1\right)}
\\
& y'=\frac{\sqrt{3} Q^2 \left(-3 c_{\theta }-2 y^2 c_{\sigma }+y^2-1\right)}{6 c_{\sigma }-3}+x^2 \left(\frac{(3 \gamma -2) Q y}{2 \sqrt{3} \left(3 c_{\theta
   }+1\right)}+\frac{1}{\sqrt{3} \left(2 c_{\sigma }-1\right)}\right)+\nonumber \\
	& -\frac{2 \sqrt{3} Q y \left(3 c_{\theta }+2 y^2 c_{\sigma }-y^2+1\right)}{9 c_{\theta
   }+3}
\\
& z'=-z \left(\frac{2 \sqrt{3} Q y^2 \left(1-2 c_{\sigma }\right)}{9 c_{\theta }+3}-\frac{Q^2 y}{\sqrt{3}}\right)+\frac{(3 \gamma -2) Q x^2 z}{2 \sqrt{3}
   \left(3 c_{\theta }+1\right)}\\
& Q'=z^2 \left(\frac{2 \sqrt{3} y^2 \left(2 c_{\sigma }-1\right)}{9 c_{\theta }+3}+\frac{Q y}{\sqrt{3}}\right)+\frac{(2-3 \gamma ) x^2 z^2}{2 \sqrt{3} \left(3
   c_{\theta }+1\right)},
\end{align}
\end{subequations}
 
The variables \eqref{KS_vars_1} are related through the constraints
\begin{subequations}\label{constraints_KS_1}
 \begin{align}
-3 c_\theta Q^2+x^2-(2 c_\sigma-1)y^2&=1,\\
Q^2+z^2&=1,
 \end{align}
\end{subequations} which are preserved by the 4D system. From the equations \eqref{constraints_KS_1} it follows that $Q$ and $z$ are bounded in the intervals $Q\in [-1, 1], \; z \in [0, 1]$ (for expanding universes $Q\geq 0$). However, since $1-2 c_\sigma$ is not necessarily non-negative it follows that $x$ and $y$ are unbounded, unless $1-2 c_\sigma\geq 0.$

The restrictions \eqref{constraints_KS_1} allow the elimination of two variables, say $x$ and $z.$ 
 This leads to the following 2-dimensional dynamical system:
\begin{subequations}
\label{KS_gen_syst}
  \begin{align}
&y'=\frac{\sqrt{3} Q^2}{3-6 c_{\sigma }}+\frac{\sqrt{3} Q y \left(c_{\theta } \left((3 \gamma -2) Q^2-4\right)+\gamma -2\right)}{6 c_{\theta
   }+2}+\frac{\sqrt{3} (\gamma -2) Q y^3 \left(2 c_{\sigma }-1\right)}{6 c_{\theta }+2}+\nonumber \\ & +\frac{1}{\sqrt{3} \left(2 c_{\sigma
   }-1\right)}-\frac{\left(Q^2-1\right) y^2}{\sqrt{3}}\\
	&Q'=\frac{\sqrt{3} (2-3 \gamma )}{18 c_{\theta }+6}+\frac{\sqrt{3} (3 \gamma -2) Q^4 c_{\theta }}{6
   c_{\theta }+2}-\frac{\sqrt{3} (3 \gamma -2) Q^2 \left(3 c_{\theta }-1\right)}{18 c_{\theta }+6}+\nonumber \\ & +\frac{\sqrt{3} (\gamma -2) \left(Q^2-1\right) y^2
   \left(2 c_{\sigma }-1\right)}{6 c_{\theta }+2}+\frac{(1-Q^2) Q y}{\sqrt{3}}
\end{align}
\end{subequations}

We shall study the general case in future work (using the $\beta-$normalization). 
Let us consider the following special case here.


\subsubsection{Special case.} 
Let us assume 
\begin{equation}
3 c_\theta\equiv c_1+3 c_2 +c_3=0
\end{equation}
(see Appendix A and  the references 
\cite{Barrow:2012qy,Sandin,Alhulaimi:2013sha}), 
and define $c^2\equiv 1-2 c_\sigma=1-2(c_1+c_3)\geq 0$.  This choice leads to a compact phase space. 

With these  special values of the $c$'s, the evolution equations for Kantowski-Sachs models simplify and
the constraint becomes
 \begin{equation}
 K+\frac{\theta^2}{3}=\hat{\mu}+3 c^2\sigma_+^2.
 \end{equation}
 The following normalized variable
 \begin{equation}
 y_1=\frac{\sqrt{3} c\sigma_+}{D}
 \end{equation}
is chosen for convenience, whence 
the variables are related through the constraints
\begin{subequations}
 \begin{align}
x^2+y_1^2=1,\\
Q^2+z^2=1.
 \end{align}
\end{subequations}
Thus, the phase space is compact with  $x\in [-1, 1], \; y_1 \in [-1, 1]$ and $Q\in [-1, 1], \; z 
\in [0, 1]$ (for expanding universes $Q\geq 0$). 

The system for $(y_1,Q)$ reduces to 

\begin{subequations}
\label{correct_KS_2}
\begin{align}
&y_1'=-\frac{\left(y_1^2-1\right) \left(3 c (\gamma -2) Q y_1+2 Q^2-2\right)}{2 \sqrt{3} c},\\
&Q'=-\frac{\left(Q^2-1\right) \left(c \left(-3 \gamma +3 (\gamma -2)
   y_1^2+2\right)+2 Q y_1\right)}{2 \sqrt{3} c}
\end{align}
\end{subequations}

Since the evolution equations are invariant under the transformation $y_1 \rightarrow -y_1$
and $c \rightarrow -c$,
without loss of generality we can assume $c>0$. Scaling the time derivative by the positive factor 
${2 \sqrt{3} c}$, we then obtain:
\begin{subequations}
\label{correct_KS_2new}
\begin{align}
&y_1'=-{\left(y_1^2-1\right) \left(3 c (\gamma -2) Q y_1+2 Q^2-2\right)},\\
&Q'=-{\left(Q^2-1\right) \left(c \left(-3 \gamma +3 (\gamma -2)
   y_1^2+2\right)+2 Q y_1\right)}
\end{align}
\end{subequations}

In tables \ref{Tab96} and \ref{Tab96b} we present the equilibrium points of the system (\ref{correct_KS_2new}) and discuss their stability.
We have that $c>0, -1 \leq Q \leq 1,  -1 \leq y_1 \leq 1$.
Some of the equilibrium
points do not exist for certain values of "c".   We have not analyzed the non-hyperbolic "stiff fluid" case, $\gamma=2$,
in which there are zero eigenvalues. Clearly, the case
$c=0$ is not included here (the GR case), 
since the equations are not valid in that case.

		
			\begin{table}[ht]
				\renewcommand{\arraystretch}{1.45}
				{%
					\begin{tabular}{|l|c|c|}\hline
						Label & Coordinates: $(y_1,Q)$	& Eigenvalues \\	\hline
						$P_{1}$ & $(0,-1)$	& $3c (2-\gamma), 2c(2-3 \gamma)$  \\[1mm]
						$P_{2}$ & $(0,1)$ & $-3c (2-\gamma), -2c(2-3 \gamma)$ \\[1mm]
						$P_{3}$ & $(-1,-1)$ & $-6c (2-\gamma), 4(1-2 c)$\\[2mm]
						$P_{4}$ & $(-1, 1)$ & $6c (2-\gamma), 4(1+2 c)$\\[2mm]
						$P_{5}$ & $(1,-1)$ & $-6c (2-\gamma),-4 (1+2 c)$\\[2mm]
						$P_{6}$ & $(1,1)$ & $6c (2-\gamma),-4(1-2 c)$\\[2mm]
						$P_{7}$ & $(-1,-2c)$ & $2(4c^2-1), 4\left[c^2 (3 \gamma -2)-1\right] $\\[2mm]
						$P_{8}$ & $(1,2c)$ & $-2(4c^2-1), -4\left[c^2 (3 \gamma -2)-1\right] $\\[2mm]
						$P_{9}$ &$\left(\frac{c (2-3 \gamma )}{d}, -\frac{2}{d}\right)$& $\frac{c \left(-e-3 \gamma +6\right)}{d},\frac{c \left(e-3 \gamma +6\right)}{d}$ \\[2mm]
						$P_{10}$ &$\left(-\frac{c (2-3 \gamma )}{d}, \frac{2}{d}\right)$& $\frac{c \left(- e +3 \gamma -6\right)}{d},\frac{c \left(e+3 \gamma -6\right)}{d}$ \\[2mm]
						\hline
					\end{tabular}
				}
				\caption{\label{Tab96} Equilibrium  points of the system \eqref{correct_KS_2new} and their eigenvalues.  We use the 
				notation $d=\sqrt{3 (\gamma -2) (3 \gamma -2) c^2+4}$ and 
				$e \equiv \sqrt{3} \sqrt{2-\gamma} \sqrt{8 c^2 (2-3 \gamma )^2-27 \gamma +22}$.}
			\end{table}
		

			\begin{table}
				\begin{center}
					\renewcommand{\arraystretch}{1.4}
						\begin{tabular}{|l|c|c|c|c|}
							\hline
							Eq. Pt.  & $\gamma$ value  & $0<c<\frac{1}{2}$ & $c=\frac{1}{2}$ & $\frac{1}{2}<c$\\
							\hline
							$P_1$ & $0\leq\gamma<\frac{2}{3}$ & $++$ source & $++$ source& $++$ source\\
							& $\gamma=\frac{2}{3}$  &$+0$ &$+0$ &$+0$\\
							& $\frac{2}{3}<\gamma<2$  & $+-$ saddle & $+-$ saddle & $+-$ saddle \\
							& $\gamma=2$  & $0-$ & $0-$ & $0-$ \\
							\hline
							$P_2$ & $0\leq\gamma<\frac{2}{3}$ &$--$sink&$--$sink&$--$sink\\
							& $\gamma=\frac{2}{3}$ & $-0$& $-0$& $-0$\\
							& $\frac{2}{3}<\gamma<2$ &$-+$ saddle&$-+$ saddle&$-+$ saddle\\
							&$\gamma=2$ & $0+$& $0+$& $0+$\\
							\hline
							$P_3$& $0\leq\gamma<2$  & $-+$ saddle & $-0$& $--$ sink\\
							& $\gamma=2$ &$0+$&$00$&$0-$\\
							\hline
							$P_4$& $0\leq\gamma<2$ & $++$ source& $++$ source & $++$ source \\
							& $\gamma=2$ &$0+$&$0+$&$0+$\\
							\hline
							$P_5$& $0\leq\gamma<2$& $--$ sink& $--$ sink& $--$ sink\\
							& $\gamma=2$  & $0-$ & $0-$ &$0+$\\
							\hline
							$P_6$& $0\leq\gamma<2$& $-+$saddle& $+0$ & $++$ source\\
							& $\gamma=2$ & $0-$ & $00$ &$0+$\\
							\hline
							$P_7$ & $0\leq\gamma<2$ & $--$ sink & $-0$ &DNE\\
							& $\gamma=2$ & $--$ sink & $00$ & DNE\\
							\hline
							$P_8$ & $0\leq\gamma<2$  & $++$ source & $+0$ &DNE\\
														&  $\gamma=2$ &$++$ source & $00$ & DNE\\
							\hline
							$P_9$ & $0\leq\gamma< \frac{2}{3}$ & $-+$ saddle & $-+$saddle & $-+$saddle\\
							& $\gamma=\frac{2}{3}$ & $+0$  & $+0$ &$+0$\\
							& $\gamma=2$ & $00$  & $00$ & $00$\\
							\hline
							$P_{10}$ & $0\leq\gamma<\frac{2}{3}$ & $-+$ saddle &  $-+$ saddle & $-+$ saddle\\
							& $\gamma=\frac{2}{3}$ & $-0$  & $-0$ & $-0$\\
							& $\gamma=2$ & $00$  & $00$ & $00$\\
							\hline
						\end{tabular}
					\end{center}
					\caption{\label{Tab96b} Stability of the equilibrium points of the system \eqref{correct_KS_2new}.}
				\end{table}

Let us enumerate the stability conditions for the hyperbolic equilibrium points:					

		\begin{enumerate}
			\item The equilibrium point $P_1$ is a source for $c>0, 0\leq \gamma <\frac{2}{3}$, and a 
			saddle  for $\frac{2}{3}<\gamma\leq 2$. Non-hyperbolic for $\gamma=\frac{2}{3}$ or 
			$\gamma=2$.  
			\item The equilibrium point $P_2$ is a sink for $c>0, 0\leq \gamma <\frac{2}{3}$, and a saddle for $\frac{2}{3}<\gamma< 2$.  Non-hyperbolic for $\gamma=\frac{2}{3}$ or $\gamma=2$.  
			\item The equilibrium point $P_3$ is a sink for $0\leq \gamma <2, c>\frac{1}{2},$ and non-hyperbolic for $c=\frac{1}{2}$ or $\gamma= 2$. Saddle otherwise. 
			\item The equilibrium point $P_4$ is a source for $c>0, 0\leq \gamma <2$. Non-hyperbolic for $\gamma=2$. 
			\item The equilibrium point $P_5$ is a sink for $c>0, 0\leq \gamma <2$. Non-hyperbolic for $\gamma=2$.
			\item The equilibrium point $P_6$ is a source for $0\leq \gamma <2, c>\frac{1}{2}.$ A saddle for $0\leq \gamma <2, 0<c<\frac{1}{2}$.  Non-hyperbolic for $\gamma=2$ or $c=\frac{1}{2}$.
			\item The equilibrium point $P_7$ exist for $0\leq \gamma \leq 2, 0<c\leq \frac{1}{2}$. It is a sink for $0\leq \gamma \leq 2, 0<c<\frac{1}{2}$. Non-hyperbolic otherwise.
			\item The equilibrium point $P_8$ exists for $0\leq \gamma \leq 2, 0<c\leq \frac{1}{2}.$ It is a source for $0\leq \gamma \leq 2, 0<c<\frac{1}{2}$. Non-hyperbolic otherwise.
			\item The equilibrium point $P_9$ exists for $0<c\leq \frac{1}{2},  0\leq \gamma \leq \frac{2}{3},$ or $0<c\leq \frac{1}{2}, \gamma=2,$ or $c>\frac{1}{2}, 0\leq \gamma \leq \frac{2}{3}.$ It is a saddle for $0\leq \gamma <\frac{2}{3}, c>0$. Non-hyperbolic for $\gamma=\frac{2}{3}$ or $\gamma=2$. 
			\item The equilibrium point $P_{10}$ exists for $0<c\leq \frac{1}{2},  0\leq \gamma \leq \frac{2}{3},$ or $0<c\leq \frac{1}{2}, \gamma=2,$ or $c>\frac{1}{2}, 0\leq \gamma \leq \frac{2}{3}.$ It is a saddle for $0\leq \gamma <\frac{2}{3}, c>0$. Non-hyperbolic for $\gamma=\frac{2}{3}$ or $\gamma=2$.
		\end{enumerate}

 {\em{Discussion.}} In the case $c_\sigma<\frac{1}{2}$ (i.e., $c>0$), when $\gamma < \frac{2}{3}$,
$P_2$ is the unique shear-free, zero curvature (FLRW) inflationary future attractor, and for
$\frac{3}{8}<c_\sigma<\frac{1}{2}$ (i.e., $0<c<\frac{1}{2}$) and $0\leq\gamma<2$ the sources and sinks are, respectively, $P_4$ \& $P_8$ and 
$P_5$ \& $P_7$. All of these sources and sinks have maximal shearing and all, except $P_7$, have zero curvature;
the sink  $P_7$ does not have zero curvature. For $c_\sigma<\frac{3}{8}$ (i.e., $c>\frac{1}{2}$) 
the points $P_7$ \& $P_8$ do not exist, and the sources and sinks with maximal shearing are $P_4$ 
and $P_5$, respectively. 

In figures \ref{fig:Syst97_gamma_0_c_0} -- \ref{fig:Syst97_gamma_1_c_0.6}
we present some orbits in the phase plane of the system 
\eqref{correct_KS_2new} for different  choices of the parameters.
In figure \ref{fig:Syst97_gamma_0_c_0}, $\gamma=0$ and $c_\sigma=0.3$. 
The sinks are $P_2, P_5$ and $P_7$. The sources are $P_1, P_4$ and $P_8$. 
$P_3, P_6, P_9$ and $P_{10}$ are saddles.
In figure \ref{fig:Syst97_gamma_0_c_0.6}, $\gamma=0$ and $c_\sigma=0.6$. The sinks are $P_2, P_3$ and $P_5$. The sources are $P_1, P_4$ and $P_6$. $P_7$ and $P_8$ do not exist. The saddles are $P_9$ and $P_{10}$. 
In figure \ref{fig:Syst97_gamma_1_c_0.2} we present the phase plane of the system \eqref{correct_KS_2new} for the choice of parameters $\gamma=1$ and $c_\sigma=0.2$. The sinks are $P_5$ and $P_7$. The sources are $P_4$ and $P_8$. The saddles are $P_1, P_2, P_3$ and $P_6$. $P_9$ and $P_{10}$ do not exist.
Finally, in figure \ref{fig:Syst97_gamma_1_c_0.6},
$\gamma=1$ and $c_\sigma=0.6$. The sinks are $P_3$ and $P_5$. The sources are $P_4$ and $P_6$. $P_1$ and $P_2$ are saddles. The points $P_7$-$P_{10}$ do not exist.

\begin{figure}
\centering
\includegraphics[width=3in, height=3in]{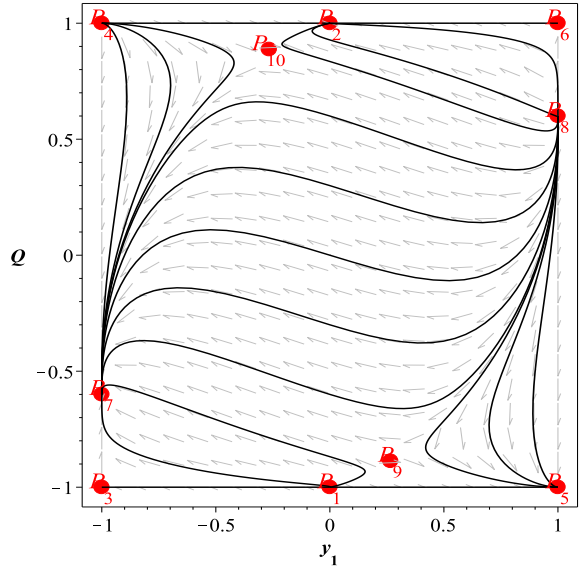}
\caption{Phase plane of the system \eqref{correct_KS_2new} for the choice of parameters $\gamma=0$ and $c_\sigma=0.3$. The sinks are $P_2, P_5$ and $P_7$. The sources are $P_1, P_4$ and $P_8$. $P_3, P_6, P_9$ and $P_{10}$ are saddles.}
\label{fig:Syst97_gamma_0_c_0}
\end{figure}
\begin{figure}
\centering
\includegraphics[width=3in, height=3in]{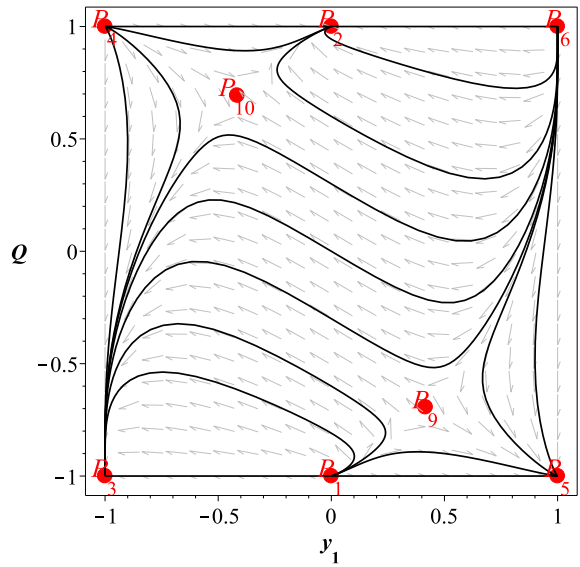}
\caption{Phase plane of the system \eqref{correct_KS_2new} for the choice of parameters $\gamma=0$ and $c_\sigma=0.6$. The sinks are $P_2, P_3$ and $P_5$. The sources are $P_1, P_4$ and $P_6$. $P_7$ and $P_8$ do not exist. The saddles are $P_9$ and $P_{10}$. }
\label{fig:Syst97_gamma_0_c_0.6}
\end{figure}

\begin{figure}[ht]
\centering
\includegraphics[width=3in, height=3in]{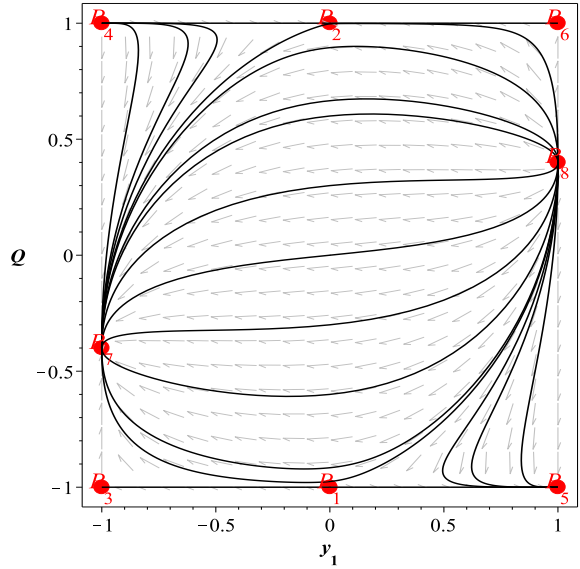}
\caption{Phase plane of the system \eqref{correct_KS_2new} for the choice of parameters $\gamma=1$ and $c_\sigma=0.2$. The sinks are $P_5$ and $P_7$. The sources are $P_4$ and $P_8$. The saddles are $P_1, P_2, P_3$ and $P_6$. $P_9$ and $P_{10}$ do not exist.}
\label{fig:Syst97_gamma_1_c_0.2}
\end{figure}

\begin{figure}
\centering
\includegraphics[width=3in, height=3in]{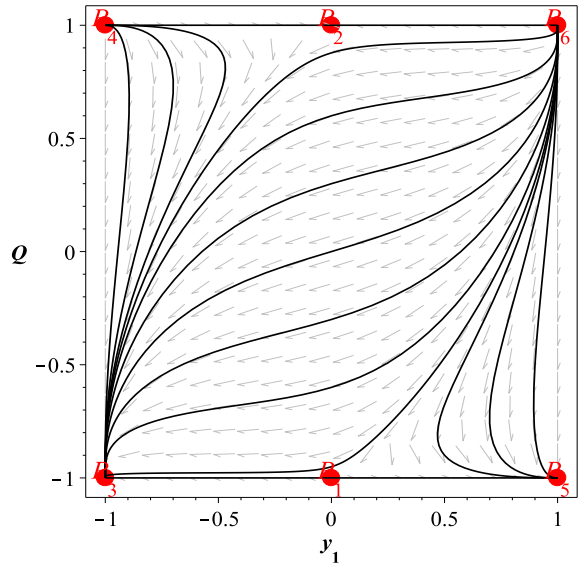}
\caption{Phase plane of the system \eqref{correct_KS_2new} for the choice of parameters $\gamma=1$ and $c_\sigma=0.6$. The sinks are $P_3$ and $P_5$. The sources are $P_4$ and $P_6$. $P_1$ and $P_2$ are saddles. The points $P_7$-$P_{10}$ do not exist.}
\label{fig:Syst97_gamma_1_c_0.6}
\end{figure}

\newpage
 
 \section{Static models}
 
 Models that are not evolving with time are also of physical importance, although
 perhaps more from the astrophysical point of view than from the cosmological one.
 In particular, much physical information can be obtained from a qualitative analysis of the models.
 Let us consider the static case $\e_0(\cdot)=0,$ for a mixture of a perfect fluid and a scalar field. 
 [In this case $\udot\equiv d\ln N/dx\neq 0$ and the perfect fluid is forced to be non-tilted ($v=0$).] We will consider barotropic equations of state $\hat{\mu}=\hat{\mu}(\hat{p})$. Since $\theta=\sp=0$ in the static subcase, $V$ depends only on the scalar field $\phi=\phi(x).$
 Furthermore, the irreducible components of the scalar field energy-momentum tensor are given 
 by $\mu^\phi=\frac{1}{2}\e_1(\phi)^2+V, p^\phi=-\frac{1}{6}\e_1(\phi)^2-V,q^\phi=0$ and 
 $\pi_+^\phi=-\frac{1}{3}\e_1(\phi)^2$ \cite{gangIIsf}. 
 
 The equations for the variables $a, \udot,\hat{p},\phi,K, N$ are:
 \begin{subequations}
 	\label{static}
 	\begin{align}
 	&\e_1 \left( a \right)= \frac{\hat{\mu }+3 \hat{p}}{2 \left(c_a+1\right)}+\e_1(\phi ){}^2-\frac{V}{c_a+1}+2 c_a \udot^2+3 a
 	\udot+K,\\
 	&\e_1 \left( \udot \right) =\frac{\hat{\mu }+3 \hat{p}}{2 \left(c_a+1\right)}-\frac{V}{c_a+1}+2
 	a \udot-\udot^2,\\
 	&{\e_1} \left( \hat{p}    \right) =- \udot(\hat{\mu}+\hat{p})\\
 	& \e_1(\e_1 (\phi))=-\left(\udot -2 a\right)\e_1 (\phi)+ V_{\phi},\\
 	&\e_1(K)= 2 a K,\label{eqKstatic}
 	\end{align}
 \end{subequations} where $V_{\phi}$ denotes differentiation with respect to $\phi.$ The system satisfies the restriction
 \begin{align}\label{static_sf}
 &a^2=c_a \udot^2+2 a \udot+\hat{p}+\frac{1}{2} \e_1(\phi ){}^2-V+K.
 \end{align}
 Taking the differential operator $\e_1(...)$ of both sides of \eqref{static_sf}, using the 
 equations \eqref{static} to substitute for the spatial derivatives, and again using the restriction 
 \eqref{static_sf} solved for $K,$ we obtain an identity. Thus, the Gauss constraint is a first integral of the system. The aether constraint is identically zero. 
 
 Let us now show how equations \eqref{static} can be used to obtain exact solutions and, additionally, use 
 the dynamical systems approach to investigate the structure of the whole solution space. 
 The first thing to do is to select a suitable  radial coordinate. We may choose
 a new radial coordinate $\lambda$ such that the equation \eqref{eqKstatic} has a trivial 
 solution. A reasonable way to do this is to select an $r$-coordinate such that $\e_1(f)\equiv -a r\partial_r 
 (f)$ (as in \cite{Clarkson:2003mp}). This implies $K\propto r^{-2}$ and $\int\ex(x)^{-1}\mathrm{d} x=-\int a^{-1} \mathrm{d}\ln r.$
 As we will see later, for the dynamical systems investigation it is better to 
 use the new time variable $\tau=\ln r$, which takes values over the whole real line.

Here we shall study the two special cases:
(i)  perfect fluid 
(previous work has assumed a comoving aether and a comoving fluid).
(ii)  vacuum (stationary with a scalar field and harmonic potential).
In particular, for the stationary aether case, it is also of interest to choose a frame in which the aether is non-comoving.
For a non-comoving stationary aether and a tilted fluid, 
it follows [in an analogous way to the static case] that the perfect fluid must be
non-tilted ($v=0$).  Additionally, since
$\theta=\sp=0$ in the stationary subcase, $V$
depends only on the scalar field $\phi=\phi(x).$
We shall present a more comprehensive analysis 
in  \cite{static}; in particular, the 
``evolution'' equations for the tilt $\alpha$,  $a$ and 
$\udot$ 
are given in the Appendix therein.

 \subsection{Static case with perfect fluid with linear equation of state and no scalar field.}
 
 To investigate this model we will use the approach of \cite{Uggla}. First, let us consider no scalar field in \eqref{static} and use the linear equation of state
 \be
 \hat{\mu}=\mu_0+(\eta-1)\hat{p},
 \ee
 where the constants $\mu_0$ and $\eta$ satisfy $\mu_0\geq 0, \eta\geq 1.$ The case $\eta=1$ corresponds to an incompressible fluid with constant energy density, while the case $\mu_0=0$ describes a scale-invariant equation of state.

 Introducing the new dimensionless variables
 \be x_1=\frac{\mu_0}{a^2}, \quad x_2=\frac{\udot}{a}, \quad
 x_3=\frac{\hat{p}}{a^2},\quad x_4=\frac{K}{a^2}, 
 \ee
 we obtain the dynamical system
 \begin{subequations}
 	\label{static2}
 	\begin{align}
 	&\frac{d x_1}{d \tau}=x_1 \left(\frac{x_1+x_3 (\eta+2)}{c_a+1}+4 x_2^2 c_a+2 x_4+6 x_2\right),\\
 	&\frac{d x_2}{d \tau}=\frac{(x_2-1) (x_1+x_3 (\eta+2))}{2 \left(c_a+1\right)}+2 x_2^3 c_a+x_2 (x_4+4 x_2-2),\\
 	&\frac{d x_3}{d \tau}=\frac{x_3 (x_1+x_3 (\eta+2))}{c_a+1}+4 x_3 x_2^2 c_a+x_2 (x_1+x_3 (\eta+6))+2 x_4
 	x_3,\\
	&\frac{d x_4}{d \tau}=x_4 \left(\frac{x_1+x_3 (\eta+2)}{c_a+1}+4 x_2^2 c_a+2 x_4+6 x_2-2\right),
 	\end{align}
 \end{subequations}
 subject to the constraint
 \begin{align}\label{static_constr}
 &1=x_2^2 c_a+2 x_2+x_4+x_3.
 \end{align}
 The constraint \eqref{static_constr} is preserved by the dynamical system  \eqref{static2}. 
 Solving the constraint \eqref{static_constr} for $x_4$ and substituting back into the system 
 \eqref{static2} we obtain the reduced system:
 
 \begin{subequations}
 	\label{syst-7}
 	\begin{align}
 	&\frac{d x_1}{d \tau}=
 	\frac{x_1 x_3 \left(\eta-2 c_a\right)}{c_a+1}+2 x_1 x_2^2 c_a+x_1
 	\left(\frac{x_1}{c_a+1}+2\right)+2 x_1 x_2,\\
 	&\frac{d x_2}{d \tau}=x_2 \left(\frac{x_1}{2 c_a+2}+x_3 \left(\frac{\eta+2}{2
 		c_a+2}-1\right)-1\right)-\frac{x_1}{2 c_a+2}-\frac{x_3 (\eta+2)}{2 c_a+2}+x_2^3 c_a+2 x_2^2,\\
 	&\frac{d x_3}{d \tau}=x_3
 	\left(\frac{x_1}{c_a+1}+2\right)+x_3^2 \left(\frac{\eta+2}{c_a+1}-2\right)+2 x_3 x_2^2 c_a+x_2
 	(x_1+x_3 (\eta+2)),
 	\end{align}
 \end{subequations}
 defined on the phase space
 \be
 \label{Static_phase_1}
 \Psi =\left\{\left(x_1, x_2, x_3\right): x_1\geq 0, x_2^2 c_a+2 x_2+x_3\leq 1\right\}.
 \ee
 
The equilibrium points of the system \eqref{syst-7} are given in table 4. Let us discuss their stability.
 		\begin{enumerate}
 			\item The equilibrium point $P_1$ is always a saddle. 
 			It satisfies $K=a^2$ asymptotically, which implies 
 			$\ex\sim e^{-\tau}=\frac{1}{r},\ey\sim e^{-\tau}=\frac{1}{r}.$ Since 
 			$\udot\ll a$ as $r\rightarrow \infty$, it follows that $N\ll r^{-1}$. 
 			\item Although the equilibrium  point $P_2$ can be an attractor for 
 			$\eta \geq 1, c_a\leq -\eta -3$ or $\eta \geq 1, -\eta -3<c_a<-1$, 
 			since it can never belong to the
 			phase space (denoted $\notin\Psi_2$ in table), we do not discuss it further. 
 			\item The equilibrium point $P_3$ is always a saddle. 
 			\item The equilibrium point $P_4$ is a source for $\eta >2, 
 			\frac{\eta -2}{4}<c_a<\frac{1}{16} (\eta -2) (\eta +6)$. Otherwise it is a saddle. 
 			\item The equilibrium point $P_5$ is a saddle for 
 			$\eta \geq 1, -1\leq c_a<-\frac{3}{4}$. It is
 		        non-hyperbolic for $c_a=-\frac{3}{4}$ [but it behaves as a saddle]. 
 			\item $P_{6}$ is a source for $\eta \geq 1,-1<c_a<0,$ or  $\eta \geq 1, c_a>0$.
 			\item $P_7$ is a sink for $1\leq \eta <2, \frac{1}{16} (\eta -2) (\eta +6)<c_a<0$. 
 			It is a source for $1\leq \eta \leq 2, c_a>0$ or $\eta >2, c_a>\frac{1}{16} (\eta -2)(\eta +6)$. It is a saddle otherwise. 
 		\end{enumerate}

 	\begin{table}
 		\renewcommand{\arraystretch}{1.45}
 		\resizebox{\columnwidth}{!}{
 		\begin{tabular}{|l|ccc|c|}\hline
 			Label		&	$x_1$	& $x_2$	& $x_3$	 &	Existence \\	\hline
 			$P_1$ 	&	0	& 0	& 0		& always 			\\[1mm]
 			$P_2$   &	$-2-\eta$	& 0	& 1	&	 $\notin \Psi$		\\[1mm]
 			$P_3$   &	0&$-\frac{2}{\eta }$&$\frac{4 \left(c_a+1\right)}{\eta ^2}$			& 	$c_a\leq \Delta_1$		\\[1mm]
 			$P_4$   &	0 &$\frac{\eta +2}{-4 c_a+\eta -2}$ &$-\frac{\left(c_a+1\right) \left((\eta -2) (\eta +6)-16
 				c_a\right)}{\left(-4 c_a+\eta -2\right){}^2}$ 	& $1\leq \eta <2, \Delta_2\leq c_a<\frac{\eta -2}{4}$ or  \\
 			&	 	&  	&  	&	  	 $\eta >2, \frac{\eta
 				-2}{4}<c_a\leq \Delta_2$	\\[1mm]
 			$P_5$  &	$-\frac{\eta  \left(c_a+1\right) \left(4 c_a+3\right)}{\left(2 c_a+1\right){}^2}$&$\frac{1}{-2
 				c_a-1}$&$\frac{\left(c_a+1\right) \left(4 c_a+3\right)}{\left(2 c_a+1\right){}^2}$  & $\eta \geq 1, -1\leq c_a\leq -\frac{3}{4}$ 	\\[1mm]
 			$P_{6,7}$ &	0 & 	$\frac{1}{1\pm \sqrt{1+c_a}}$ & 0  & always	\\[2mm]\hline
 		\end{tabular}
 		}
 		\caption{Equilibrium  points of the system \eqref{syst-7}. We use the notation $\Delta_1=\frac{1}{8} \left(\eta ^2+4 \eta -4\right)$ and $\Delta_2=\frac{1}{16} \left(\eta ^2+4 \eta -12\right).$}
 	\end{table}

 \begin{table}
 	\renewcommand{\arraystretch}{1.45}
 	\centering
 		\begin{tabular}{|l|c|}\hline
 			Label	& Eigenvalues	\\	\hline
 			$P_1$ 	& $-1,2,2$\\[1mm]
 			$P_2$   & $-2,-1+ \frac{c_a+\eta +3}{\sqrt{\left(c_a+1\right) \left(c_a+\eta +3\right)}},-1- \frac{c_a+\eta +3}{\sqrt{\left(c_a+1\right) \left(c_a+\eta +3\right)}}$\\[1mm]
 			$P_3$   & $2,-\frac{\eta +2+ \sqrt{64 c_a-7 \eta  (\eta +4)+36}}{2 \eta },-\frac{\eta +2- \sqrt{64 c_a-7 \eta  (\eta +4)+36}}{2 \eta }$\\[1mm]
 			$P_4$   & $\frac{\eta ^2-4}{2 \left(4 c_a-\eta +2\right)}-2,\frac{\eta  (\eta +2)}{4 c_a-\eta +2}-2,\frac{\eta  (\eta +2)}{4 c_a-\eta
 				+2}$\\[1mm]
 			$P_5$    & $-\frac{\eta }{2 c_a+1},-2,\frac{1}{-2 c_a-1}-2$\\[1mm]
 			$P_{6,7}$& $\frac{2 \left(\left(c_a\mp \sqrt{c_a+1}\right)+1\right)}{c_a},\frac{(\eta +6) \left(1\pm \sqrt{c_a+1}\right)+4
 				c_a}{\left(\sqrt{c_a+1}-1\right){}^2},\frac{2 \left(\left(2 c_a\mp \sqrt{c_a+1}\right)+1\right)}{c_a}$\\[1mm]\hline
 		\end{tabular}
 	\caption{Eigenvalues of the equilibrium  points of the system \eqref{syst-7}.}
 \end{table}

 \subsection{Static vacuum aether with a scalar field with harmonic potential.}
 
 Let us investigate a static vacuum aether with a scalar field with harmonic potential 
 $V(\phi)=\frac{m^2\phi^2}{2}$  (also see \cite{gangIIsf}).
 Introducing the new dimensionless variables
 \be  \quad x_2=\frac{\udot}{a}, \quad
 x_4=\frac{K}{a^2}, \quad x_5=\frac{\sqrt{2}m}{a}, \quad
 x_6=\frac{\e_1(\phi)}{a}, \quad x_7=\frac{m \phi}{\sqrt{2} a},
 \ee
 we obtain the dynamical system 
 \begin{subequations}\label{static3}
 	\begin{align}
 	&\frac{d x_2}{d \tau}=2 x_2^3 c_a-\frac{(x_2-1) x_7^2}{c_a+1}+x_2 \left(x_4+4 x_2+x_6^2-2\right),\\
 	&\frac{d x_4}{d \tau}=2 x_4 \left(2 x_2^2 c_a-\frac{x_7^2}{c_a+1}+x_4+3 x_2+x_6^2-1\right),\\
 	&\frac{d x_5}{d \tau}=x_5 \left(2 x_2^2 c_a-\frac{x_7^2}{c_a+1}+x_4+3 x_2+x_6^2\right),\\
	&\frac{d x_6}{d \tau}=x_6 \left(-\frac{x_7^2}{c_a+1}+x_4+4 x_2+x_6^2-2\right)+2 x_2^2 x_6 c_a-x_5 x_7,\\
 	&\frac{d x_7}{d \tau}=x_7 \left(-\frac{x_7^2}{c_a+1}+x_4+3 x_2+x_6^2\right)+2 x_2^2 x_7 c_a-\frac{x_5 x_6}{2},
 	\end{align}
 \end{subequations}
 subject to the constraint
 \begin{align}\label{static_constr_sf}
 &1=c_a {x_2}^{2}+ 2 x_2+ x_4+\frac{1}{2}x_6^2-x_7^2.
 \end{align}
 The constraint \eqref{static_constr_sf} is preserved by the dynamical system  \eqref{static3}. 
 Solving the constraint \eqref{static_constr_sf} for $x_4$ and substituting back 
 into the system \eqref{static3} we obtain the reduced system:
 \begin{subequations}\label{static4}
 	\begin{align}
 	&\frac{d x_2}{d \tau}=x_2^3
 	c_a-\frac{(x_2-1) x_7^2}{c_a+1}+\frac{1}{2} x_2 \left(4 x_2+2 x_7^2+x_6^2-2\right),\\
 	&\frac{d x_5}{d \tau}=\frac{1}{2} x_5 \left(2 c_a \left(\frac{x_7^2}{c_a+1}+x_2^2\right)+2 x_2+x_6^2+2\right),\\
	&\frac{d x_6}{d \tau}=x_6 c_a
 	\left(\frac{x_7^2}{c_a+1}+x_2^2\right)+\frac{1}{2} \left(x_6 \left(4 x_2+x_6^2-2\right)-2 x_5
 	x_7\right),\\
 	&\frac{d x_7}{d \tau}=x_7 c_a \left(\frac{x_7^2}{c_a+1}+x_2^2\right)-\frac{x_5 x_6}{2}+x_7
 	\left(x_2+\frac{x_6^2}{2}+1\right),
 	\end{align}
 \end{subequations}
 defined in the phase space
 \be
  \label{Static_phase_2}
 \Psi =\left\{\left( x_2, x_5, x_6, x_7\right): c_a {x_2}^{2}+ 2 x_2+ x_4+\frac{1}{2}x_6^2-x_7^2 \leq 1\right\}.
 \ee

 \begin{table}
 	\renewcommand{\arraystretch}{1.45}
 	\resizebox{\columnwidth}{!}
 	{
 		\begin{tabular}{|l|cccc|c|}\hline
 			Label		& $x_2$	&	$x_5$	& $x_6$	&	$x_7$ &	Existence \\	\hline
 			$Q_{1}$ 	&	 0 & 0	&  0	&	 0	& always 			\\[1mm]
 			$Q_{2,3}$ 	&	 ${x_2^\star}$ & 0	&  $\pm \sqrt{2} \sqrt{1-2 {x_2^\star-c_a{x_2^\star}^2}}$	 	&	 0	& $1-2 {x_2^\star-c_a{x_2^\star}^2}\geq 0$ 			\\[1mm]
 			$Q_{4,5}$  &$-\frac{1}{2 c_a+1}$ &   0 & 0 &$\pm\frac{\sqrt{\left(-c_a-1\right) \left(4 c_a+3\right)}}{2
 				c_a+1}$ & $\left(c_a+1\right) \left(4 c_a+3\right)\leq 0$ \\[1mm]
 			$Q_{6,7}$  &$-\frac{1}{2 c_a+1}$ &  0 &$\pm\frac{\sqrt{2} \sqrt{\left(c_a+1\right) \left(4 c_a+3\right)}}{2
 				c_a+1}$ & 0  & $\left(c_a+1\right) \left(4 c_a+3\right)\geq 0$ \\[1mm]
 			$Q_{8,9}$ &$\frac{1}{1\pm \sqrt{c_a+1}}$ &  0  & 0 & 0  & always \\[1mm]\hline
 		\end{tabular}
 	}
	 	\caption{Equilibrium  points of the system \eqref{static4}. $x_2^\star$ is a parameter and 
	 	hence the curves $Q_{2,3}$ represent lines of equilibrium  points
               ($x_2^\star=0, x_2^\star=2$ are special points on these curves).}
 \end{table}
 
 \begin{table}
 	\renewcommand{\arraystretch}{1.45}
 	\centering
 		\begin{tabular}{|l|c|}\hline
 			Label	& Eigenvalues 	\\	\hline
 			$Q_{1}$ & $-1,-1,1,1$	\\[1mm]
 			$Q_{2,3}$   & $2-x_2^\star, 2-x_2^\star, 2(1-x_2^\star),0$  \\[1mm]
 			$Q_{4,5}$   & $-\frac{4 c_a+3}{2 c_a+1},-\frac{4 c_a+3}{2 c_a+1},-2,0$\\[1mm]
 			$Q_{6,7}$   & $\frac{4 \left(c_a+1\right)}{2 c_a+1},\frac{4 c_a+3}{2 c_a+1},\frac{4 c_a+3}{2 c_a+1},0$ \\[1mm]
 			$Q_{8,9}$    & $\frac{\left(2 c_a\mp \sqrt{c_a+1}\right)+1}{c_a},\frac{\left(2 c_a\mp \sqrt{c_a+1}\right)+1}{c_a},\frac{2 \left(\left(c_a\mp
 				\sqrt{c_a+1}\right)+1\right)}{c_a},0$ \\[2mm]
 			\hline
 		\end{tabular}
		\caption{Eigenvalues of the equilibrium  points of the system \eqref{static4}
		given in the previous table.}
  \end{table}

The equilibrium points of the system \eqref{static4} are described in tables 6 and 7.
Let us discuss their stability. 		
 		\begin{enumerate}
 			\item $Q_1$ is always a saddle. 
 			\item The line of equilibrium points $Q_{2,3}$ is normally hyperbolic and is stable when $x_2^\star>2$.
 			\item The equilibrium  points $Q_{4,5}$ are non-hyperbolic. 
 			They have a 3D stable manifold and a 1D center manifold for 
 			$-1\leq c_a<-\frac{3}{4}$ and a 1D stable manifold and a 3D center manifold for 
 			$c_a=-\frac{3}{4}$. 
 			\item $Q_{6,7}$ are non-hyperbolic. They have a 3D stable manifold and 
 			a 1D center manifold for $-\frac{3}{4}<c_a<-\frac{1}{2}.$ They have a 
 			3D unstable manifold and a 1D center manifold for $c_a<-1$ or $c_a>-\frac{1}{2}$ [the non zero eigenvalues are always of the same sign].
 			\item $Q_8$ is non-hyperbolic. It has a 3D unstable manifold and a 1D center manifold for $-1<c_a<0$ or $c_a>0$. Otherwise, its center manifold has dimension greater than 1. 
 			\item $Q_9$ is non-hyperbolic. It has a 3D stable manifold and a 
 			1D center manifold for $-\frac{3}{4}<c_a<0.$ It has a 3D unstable manifold and a 
 			1D center manifold for $c_a>0$. Finally, $Q_9$ has a 2D unstable manifold, 
 			a 1D center manifold and a 1D stable manifold for $-1<c_a<-\frac{3}{4}.$
 		\end{enumerate}

Although static models are of particular physical importance, in this paper we have primarily focused
on the mathematical properties of the solution space.
It can be observed that the phase spaces \eqref{Static_phase_1} and
\eqref{Static_phase_2} are in general non-compact; thus a more detailed analysis requires the introduction of
compact variables.  In addition, since the equilibrium points in
\eqref{Static_phase_2} are generically non-hyperbolic, the use of the center manifold theorem is required, 
which is beyond the linear analysis provided here.  A more detailed stability analysis for the equilibrium
points of both the dynamical systems \eqref{syst-7} and \eqref{static4}, and the study of the tilted aether
static model, is left for the companion paper
\cite{static}.


 \newpage
 \section{Discussion}

In this paper we have  studied spherically symmetric Einstein-aether models 
with tilting perfect fluid matter, which are also solutions of the IR limit of
Horava gravity \cite{TJab13}. 
We used the 1+3 frame formalism \cite{EU,WE,SSSS} to write down the
evolution equations for non-comoving perfect fluid spherically symmetric models and showed they form a
well-posed system of first order PDEs in two variables.
We adopted the so-called
comoving aether gauge (which implies a preferred foliation,
the only remaining freedom is the coordinate time and space reparameterization freedom).
We also introduced  ($\beta$-) normalized variables. 
The formalism is particularly well-suited for numerical and qualitative analysis. 
In particular,
we considered the special subset  $\dot{U}=v=0$
(where we also assumed   $c_\theta=0$ and $c_\sigma \neq 0$) and derived
the final reduced phase space equations in normalized variables.

The formalism adopted here is 
appropriate for the
study of the qualitative properties of astrophysical and cosmological models with 
values for the non-GR parameters which are consistent with current constraints.
In particular, motivated by current cosmological observations, we have
studied inhomogeneous cosmologies in Einstein-aether theories of gravity.

We first considered dust models. We investigated a special dust model with $\dot{U}=0$ and $v = 0$ in normalized variables
(assuming
$c_{\sigma } \neq 0$) and derived a reduced (closed) evolution system.
The FLRW models in this special dust model correspond to an equilibrium point.
In these models we are particularly interested 
physically in their late time evolution.
Therefore, we paid particular attention to the sinks for different values of the parameter $ c_{\sigma }$
(which were summarized earlier).
In all cases $\Omega \rightarrow 0$
to the future. For all solutions with  small  ${c_\sigma }< 3/8$, $\mathcal{D} \rightarrow 1$ 
($\mathcal{Q} \rightarrow 1$) and the shear goes to zero at late times. Consequently, the models close
to GR isotropize to the future.

We briefly reviewed the FLRW models in which the source must be of the form of a comoving perfect fluid (or vacuum)
and the aether must be comoving.
We then considered the spatially homogeneous 
Kantowski-Sachs models \cite{kramer} using appropriate 
normalized variables (which are bounded; note that we did not use 
the $\beta-$normalization here), and obtained the general evolution equations.
We then considered a special case with $3 c_\theta\equiv c_1+3 c_2 +c_3=0$ and
analysed the qualitative behaviour. 
In this case a full global dynamical analysis
is possible, and we determined both the  
early and late  time behaviour of the models and their physical properties.

In the case $c_\sigma<\frac{1}{2}$ (i.e., $c>0$), when $\gamma < \frac{2}{3}$, there is the unique shear-free,
zero curvature (FLRW) inflationary future attractor ($P_2$), and for $\frac{3}{8}<c_\sigma<\frac{1}{2}$ (i.e.,
$0<c<\frac{1}{2}$) and $0\leq\gamma<2$ all of the sources and sinks (respectively, $P_4$ \& $P_8$ and $P_5$ \&
$P_7$) have maximal shearing and all except one sink ($P_7$) have zero curvature.  For $c_\sigma<\frac{3}{8}$
(i.e., $c>\frac{1}{2}$), the points $P_7$ \& $P_8$ do not exist, and the sources and sinks with maximal shearing
are $P_4$ and $P_5$, respectively.

Finally, we considered static models  for a mixture of a (necessarily non-tilted with $v=0$) perfect fluid with a barotropic equations of state
and a scalar field
(with a self-interaction potential $V$ that depends only on the scalar field).
In particular, we studied the special cases of a
tilted perfect fluid and no scalar field
(previous work had assumed a comoving aether and a comoving fluid) and
a stationary  vacuum with a scalar field (with a harmonic potential).
The equilibrium points in the resulting dynamical systems in these two cases 
were determined and their stability was investigated. 
Although models that are not evolving with time are of physical importance, 
and physical information can be obtained from their qualitative analysis,
we have primarily focussed
on the mathematical properties of the solution space in this paper.
The physical interpretation of this analysis 
will be comprehensively discussed in \cite{static}.

We also examined the conditions for the existence of McVittie-like
solutions in the context of Einstein-aether theory.  We found that they only
exist for the choice of
parameters $c_a=0,\gamma=0$, and for an aligned aether ($v=0$).  Since $\gamma=0$, the matter fluid corresponds
to a cosmological constant (and $\theta$ is
always a constant).
Irrespective of the sign of the initial expansion, the
physical variables tend to zero as $t\rightarrow +\infty$.

In future work we shall investigate the general Kantowski-Sachs models and the static models
more comprehensively. 
In particular, it would be
of interest to determine the structure of stationary rotating solutions;
rapidly rotating black holes,
unlike the non-rotating ones, might turn out to be
very different from the Kerr metrics of GR.

We note that the tilt is defined relative to matter; one important question is to investigate whether this tilt decays
to the future in general.
We shall also study 
spherically symmetric, self-similar spacetimes which also admit, in addition to the three Killing vectors, 
a homothetic vector  \cite{SSSS}.
 
 \newpage
 
 \appendix
 
 \section{Models and the parameters $c_i$} 
 
 We can study different models with different dimensionless parameters $c_i$.
 From earlier:
 \begin{displaymath}
 c_\theta = c_2 + (c_1 + c_3)/3,\ c_\sigma = c_1 + c_3,\ c_\omega = c_1 - c_3,\ c_a = c_4 - c_1.
  \end{displaymath} 
 Since the spherically symmetric models are
 hypersurface orthogonal the aether field has vanishing twist and the field equations are
 therefore independent
 of the twist parameter $c_\omega$ \cite{TJab13} (this is equivalent to
 being able to set $c_4=0$ without loss of
 generality \cite{Jacobson}).

 A second condition on the $c_i$ can effectively be specified
 by a renormalization the Newtonian
 gravitational constant $G$. From \cite{Jacobson} we have that $G_N = G\left(1 - 
 \frac{1}{2}(c_1 + c_4)\right)^{-1}$. So long as $(c_1 + c_4) <2$, so that
 the gravitational constant is positive, we can effectively renormalize and specify
 $(c_1 + c_4)$. If not, and we reduce the theory to a one parameter model, the theory might be pure GR in disguise
 [in GR $c_i=0$].

 The remaining two non-trivial constant parameters in the model 
 must satisfy additional constraints
 (it will be useful here to define
 $c^2 \equiv 1 - 2c_\sigma \leq 1$):

 \paragraph{Observations:}
 
 The models (i.e., the values of the $c_i$) must be consistent with all observations.
 In general, if the magnitudes of all of the $c_i$ are 
 (non-zero and) small (e.g., less than $10^{-2}$), then the models will be physical 
 \cite{Jacobson,Barausse:2011pu}.

 \paragraph{Self-consistency:}
 
 There are also a number of self-consistency requirements \cite{Jacobson,Barausse:2011pu}:
 
 \be
 \label{sc1}
 0 \leq c_1 + c_3  \leq 1,
 \ee
 \be
 \label{sc2}
 0 \leq c_1 - c_3  \leq     \frac{(c_1 + c_3)}{3[1- (c_1 + c_3)]},
 \ee
 which can be written in terms of 
 $c_\theta,\ c_\sigma ,\ c_a$. Note that this imples that $0 \leq c_1$,  $c_3 \leq c_1 \leq  1-c_3$.

 \subsection{Case A:}
 
 All of the $c_i$ are small (and not all zero). We set $c_4=0$. We renormalize and chose $c_1$
 so that $c_1+3c_2+c_3=0$ (i.e., $c_\theta=0$). The self-consistency relations (\ref{sc1},  \ref{sc2})
 then imply that  $0 \leq c_3 \leq c_1 \leq  2c_3$. We thus have a two parameter model with small
 $c_\sigma \geq 0,\ c_a \leq 0$. Using  $c_1 \equiv d c_3$
 ($1 \leq d \leq 2$), we have that $c_\sigma=\frac{1}{2}(1- c^2)$ and $c_a = -\frac{d}{(1+d)} c_\sigma$.
 Note that it is not possible for $c_a = -c_\sigma$ in this case.

 {\bf{Summary case A: $c_\sigma=\frac{1}{2}(1- c^2)  \geq 0, 
 		c_a =  -\frac{d}{(1+d)} c_\sigma \leq 0, c_\theta=0$ }}

 \subsection{Case B:}
 
 We set 
 \be
 \label{PPN1}
 c_4 = -\frac{c_3{^2}}{c_1},
 \ee
 \be
 \label{PPN2}
 c_2 = -\frac{(2c_1{^2} + c_1 c_3 - c_3{^2})}{3c_1}
 \ee
 (before the field redefinition of $c_4$), so that the parameterized post-Newtonian (PPN) parameters 
 $\alpha_1=\alpha_2=0$ (and hence all solar system tests are trivially satified  \cite{Jacobson}).
 A two parameter family of models ($c_1 \neq 0, c_3$) satisfying the consistency conditions
 (\ref{sc1},  \ref{sc2}) results. Here ({\bf{Summary}}):

 \be
 c_a = -\frac{(c_1{^2} +  c_3{^2})}{c_1} \leq 0,
 \ee
 \be
 0 \leq c_\sigma  = {c_1} + {c_3} \leq 1,
 \ee
 \be
 c_\theta = -\frac{(c_1{^2} -  c_3{^2})}{3c_1} \leq 0,
 \ee
 and the $c_i$ satisfy 
 
 \be
 (c_a{^2} - 9c_\theta{^2}) = \left(c_a + 3c_\theta +2 c_\sigma\right)^2.
 \ee

 \subsubsection{Case B(ii):}
 
 If we also renormalise the Newtonian gravitational potential by setting
 $c_1+c_4=0$, then from (\ref{PPN1}) we obtain $c_1{^2}=c_3{^2}$, and hence 
 $c_1=c_3$ (since $c_1=-c_3$ cannot satisfy (\ref{sc1},  \ref{sc2})). In this case
 $c_\theta=0$. Hence, $c_a = - c_\sigma$, and we have that 
 $c_\sigma=\frac{1}{2}(1- c^2)  \geq 0$ and $c_a=-\frac{1}{2}(1- c^2)  \leq 0$,
 where $c^2 \leq 1$ but need not be small (and the self-consistency relations
 (\ref{sc1},  \ref{sc2}) are satisfied).

 {\bf{Summary case B(ii): $ c_\sigma=\frac{1}{2}(1- c^2)  \geq 0,  c_a=-\frac{1}{2}(1- c^2),  
 		c_\theta=0$ }}
 
 \subsection{Case C:}
 
 In principle we can study the physics of the models for different parameter ranges
 of the $c_i$. If we study the models in the early universe (where the constants $c_i$
 can be replaced with evolving parameters \cite{kann}), then the observational constraints 
 above need not apply.
 
 In one particularly interesting theoretical case 
 (see section 4), we could consider $3c_\theta + 2c_\sigma=0$ 
 [we could also use the renormalization of the Newtonian
 gravitational constant and consider the case $c_a =0$].
 Note that $3c_\theta + 2c_\sigma=c_1+c_2+c_3=0$. This implies that the PPN parameter $\alpha_1$
 diverges \cite{Jacobson} and conditions (\ref{sc1},  \ref{sc2}) can only be satisfied when
 $c_1=c_2=c_3=0$, the GR case. However, for theoretical reasons it may be of interest to study
 this case in early universe cosmological models.

 {\bf{Summary case C: $c_\sigma=\frac{1}{2}(1- c^2)  \geq 0, 
 		c_\theta = -\frac{1}{3}(1- c^2) \leq 0, [c_a = 0]$ }}
 
 \newpage

  \section{Further development of the governing equations}\label{App3} 
Let us further develop the governing equations presented in Section 3.
Combining equations \eqref{eq99} and \eqref{eq100} and using the identity $\Sigma=1-\mathcal{Q},$ we obtain 
\begin{align}
& q= -\frac{3 \mathcal{K}}{2(2 c_{\sigma }-1)}+\frac{1}{3} \mathcal{A}\Udot \left(\frac{2 \left(c_a+1\right)}{3 c_{\theta }+1}+\frac{1-2 c_a}{1-2
   c_{\sigma }}\right)-\frac{\left(c_a+1\right) \left(3 c_{\theta }+2 c_{\sigma }\right) \pmb{\partial}_1 \Udot}{3 \left(3 c_{\theta }+1\right) \left(2
   c_{\sigma }-1\right)}+\nonumber \\ & +\frac{1}{3} r \left(c_a+1\right) \left(\frac{1}{3 c_{\theta }+1}+\frac{1}{2 c_{\sigma }-1}\right) \Udot+\frac{1}{6}
   \Udot^2 \left(\frac{5 c_a+2}{1-2 c_{\sigma }}-\frac{2 \left(c_a+1\right)}{3 c_{\theta }+1}\right)+\nonumber\\
	& +\frac{\mathcal{A}^2}{2(2 c_{\sigma }-1)}+\frac{4\left(3
   c_{\theta }+2 c_{\sigma }\right) \mathcal{Q}}{3 c_{\theta }+1}+\left(\frac{6-3 c_{\theta }-14 c_{\sigma }}{4 c_{\sigma }-2}+\frac{2-4 c_{\sigma }}{3
   c_{\theta }+1}+1\right) \mathcal{Q}^2+\nonumber \\ 
	& -\frac{\Omega \left(v^2 \left(3 (\gamma +1) c_{\theta }+2 (\gamma -2) c_{\sigma }+3\right)-6
   \gamma  c_{\sigma }-3 c_{\theta }+4 c_{\sigma }+3 \gamma -3\right)}{2 \left(3 c_{\theta }+1\right) \left(2 c_{\sigma }-1\right)
   \left(1-v^2\right)}+\nonumber \\ & +\frac{1-9 c_{\theta }-8 c_{\sigma }}{6 c_{\theta }+2}. \label{def_q}
\end{align}
Combining equations \eqref{eq98}, \eqref{eq101}  and \eqref{eq106} and \eqref{eq107} and using the identity $\Sigma=1-\mathcal{Q},$ we obtain
\begin{subequations}
\begin{align}
&r=(q-1) c_a \Udot -c_a \pmb{\partial}_0 \Udot - 3 \mathcal{A} (1- \mathcal{Q})+\frac{3 \gamma  v \Omega}{2(1-v^2)}, \label{def_r}\\
& \pmb{\partial}_0 \mathcal{A}  = (q + 3(1-\mathcal{Q}))\mathcal{A} - \Udot + r,\\
&\pmb{\partial}_1\mathcal{Q}= -\frac{3 \left(2 c_{\sigma }-1\right) \mathcal{A} (1-\mathcal{Q})}{3 c_{\theta }+2 c_{\sigma }}+r \left(\mathcal{Q}+\frac{1-2 c_{\sigma }}{3
   c_{\theta }+2 c_{\sigma }}\right) -\frac{3 \gamma  v \Omega}{2 \left(3 c_{\theta }+2 c_{\sigma }\right) \left(1-v^2\right)},\label{spatial_Q}.   
\end{align}
\end{subequations}
Finally, solving the equations \eqref{def_q} and \eqref{def_r} for $q$ and $r$ we obtain:
\begin{subequations}
\label{q_r_defs}
\begin{align}
& q=-\frac{1}{{3 \left(2-\frac{2 c_a \left(c_a+1\right) \left(3 c_{\theta }+2
   c_{\sigma }\right)\Udot^2}{3 \left(3 c_{\theta }+1\right) \left(2 c_{\sigma }-1\right)}\right)}}\Big\{\frac{3 \mathcal{A}^2}{1-2 c_{\sigma }}-\frac{4 \left(c_a+1\right) \Udot\mathcal{A}}{3 c_{\theta }+1}+\nonumber \\ & +\frac{2 \left(1-2 c_a\right) \Udot
   \mathcal{A}}{2 c_{\sigma }-1}-6 \mathcal{Q}^2+\frac{2 \left(c_a+1\right) \Udot^2}{3 c_{\theta }+1}+\frac{\left(5 c_a+2\right) \Udot^2}{2 c_{\sigma
   }-1}+\nonumber \\ & +\frac{12 (\mathcal{Q}-1)^2 \left(2 c_{\sigma }-1\right)}{3 c_{\theta }+1}+\frac{3 \mathcal{Q} \left(-16 c_{\sigma }+\mathcal{Q} \left(3 c_{\theta }+14 c_{\sigma
   }-6\right)+8\right)}{2 c_{\sigma }-1}+\nonumber \\ & -\frac{3 \left((\gamma -2) v^2-3 \gamma +2\right) \Omega}{\left(3 c_{\theta }+1\right)
   \left(v^2-1\right)} -\frac{3 \left((\gamma +1) v^2-1\right) \Omega}{\left(2 c_{\sigma }-1\right)
   \left(v^2-1\right)}+\nonumber \\ & +\frac{\left(c_a+1\right) \left(3 c_{\theta }+2 c_{\sigma }\right) \Udot \left(-6 \mathcal{A} (\mathcal{Q}-1)+2 c_a
   (\pmb{\partial}_0\Udot+\Udot)+\frac{3 \gamma  v\Omega}{v^2-1}\right)}{\left(3 c_{\theta }+1\right) \left(2 c_{\sigma
   }-1\right)}+\nonumber \\ & +\frac{2 \pmb{\partial}_1\Udot \left(c_a+1\right)}{3 c_{\theta }+1}+\frac{3}{1-2 c_{\sigma }}+\frac{9 \mathcal{K}}{2 c_{\sigma }-1} +\frac{2
   \pmb{\partial}_1\Udot \left(c_a+1\right)}{2 c_{\sigma }-1}+\frac{3}{2 c_{\sigma }-1}+9\Big\},
\end{align}
\begin{align}
	& r= 3 \mathcal{Q} \mathcal{A}-3
   \mathcal{A}-c_a \pmb{\partial}_0\Udot -c_a \Udot+\frac{3 \gamma  v \Omega}{2(1- v^2)}+\nonumber \\ & -\frac{c_a \Udot }{3 \left(2-\frac{2 c_a \left(c_a+1\right) \left(3 c_{\theta }+2 c_{\sigma }\right) \Udot^2}{3 \left(3 c_{\theta }+1\right) \left(2
   c_{\sigma }-1\right)}\right)} \Big\{\frac{3 \mathcal{A}^2}{1-2
   c_{\sigma }}-\frac{4 \left(c_a+1\right) \Udot \mathcal{A}}{3 c_{\theta }+1}+\nonumber \\ & +\frac{2 \left(1-2 c_a\right) \Udot \mathcal{A}}{2 c_{\sigma }-1}-6 \mathcal{Q}^2+\frac{2
   \left(c_a+1\right) \Udot^2}{3 c_{\theta }+1}+\frac{\left(5 c_a+2\right) \Udot^2}{2 c_{\sigma }-1}+\nonumber \\ & +\frac{12 (\mathcal{Q}-1)^2 \left(2 c_{\sigma
   }-1\right)}{3 c_{\theta }+1}+\frac{3 \mathcal{Q} \left(-16 c_{\sigma }+\mathcal{Q} \left(3 c_{\theta }+14 c_{\sigma }-6\right)+8\right)}{2 c_{\sigma }-1}+\nonumber \\ & -\frac{3
   \left((\gamma -2) v^2-3 \gamma +2\right) \Omega}{\left(3 c_{\theta }+1\right) \left(v^2-1\right)}-\frac{3 \left((\gamma +1)
   v^2-1\right) \Omega}{\left(2 c_{\sigma }-1\right) \left(v^2-1\right)}+\nonumber \\ & +\frac{\left(c_a+1\right) \left(3 c_{\theta }+2
   c_{\sigma }\right) \Udot \left(-6 \mathcal{A} (\mathcal{Q}-1)+2 c_a (\pmb{\partial}_0\Udot+\Udot)+\frac{3 \gamma  v \Omega}{v^2-1}\right)}{\left(3 c_{\theta }+1\right) \left(2 c_{\sigma }-1\right)}+\nonumber \\ & +\frac{2 \pmb{\partial}_1\Udot \left(c_a+1\right)}{3 c_{\theta
   }+1}+\frac{3}{1-2 c_{\sigma }}+\frac{9 \mathcal{K}}{2 c_{\sigma }-1}+\frac{2 \pmb{\partial}_1\Udot \left(c_a+1\right)}{2 c_{\sigma }-1}+\frac{3}{2 c_{\sigma
   }-1}+9\Big\}.
\end{align}
\end{subequations} 

These equations give the expressions for $q$ and $r$ in terms of the normalized variables and the 
derivatives $\pmb{\partial}_0\Udot,
\pmb{\partial}_1\Udot.$ The expression \eqref{spatial_Q}, is used to eliminate the spatial derivative 
$\pmb{\partial}_1\mathcal{Q}$ from the equations.


The final equations for the reduced phase space $\left(\mathcal{N}, \Udot, E_1^1, \mathcal{K}, \mathcal{Q}, \Omega, v\right)^T$ are equations
(\ref{eq96},\ref{eq97},\ref{eq99},\ref{eq102},\ref{eq103}) and (\ref{eq107}) for $\pmb{\partial}_0 \mathcal{A}$, subject to the constraints
(\ref{eq104}-\ref{eq106}) and a constraint for $\pmb{\partial}_1\mathcal{Q}$ (rather than $\pmb{\partial}_1{\Sigma}$), with $q$ and $r$ defined
as in \eqref{q_r_defs}.  From the equations we have either the evolution equation for 
$\Udot$ given by \eqref{eq98} or the definition of $r$ given by
\eqref{def_r} (or by \eqref{q_r_defs}(b), after $q$-elimination in \eqref{def_r}).

The commutator equation \eqref{comm} can be expressed in terms of the normalized variables by 
\begin{align}
&[\pmb{\partial}_0,\pmb{\partial}_1]=\frac{1}{\beta^2} [\e_0,\e_1]+(1+q)\pmb{\partial}_1-r\pmb{\partial}_0\nonumber\\
& =(\Udot -r)\pmb{\partial}_0+(q+3-3\mathcal{Q})\pmb{\partial}_1.
\end{align}
On the other hand 
\begin{align}
&\pmb{\partial}_0 r= \pmb{\partial}_0\pmb{\partial}_1(-\ln \beta)
=-[\pmb{\partial}_0,\pmb{\partial}_1]\ln\beta-\pmb{\partial}_1\pmb{\partial}_0 \ln\beta\nonumber\\
&=\Udot(1+q)+r(2-3\mathcal{Q})+\pmb{\partial}_1 q, 
\end{align}                                                                                 
where we have used $\pmb{\partial}_0 \ln\beta=-(1+q)$ and $q$ is given by \eqref{def_q}.
{\footnote{ Note that the evolution equation for $r$ will contain the second order derivative term         
$\parb_1\parb_1 \Udot$ (via the term $\pmb{\partial}_1 q$), as occurs in the GR setting       
\cite{WainwrightLim}. }}

The final equations for the reduced phase space 
$\left(\mathcal{N}, \Udot, E_1^1, \mathcal{K}, \mathcal{Q}, \Omega, v, r\right)^T$ are then
\begin{subequations}
\label{reduced_syst_2:evol}
\begin{align}
& \pmb{\partial}_0 E_1^1  = (q + 3\Sigma) E_{1}^1,\\
&  \pmb{\partial}_0 \mathcal{K}  = 2q \mathcal{K},\\
&c_a \pmb{\partial}_0 \Udot=(q-1) c_a \Udot -r - 3 \mathcal{A} (1- \mathcal{Q})+\frac{3 \gamma  v \Omega}{2(1-v^2)}\label{50.c},\\
&  \pmb{\partial}_0 \mathcal{Q}-\frac{\left(c_a+1\right) \pmb{\partial}_1\Udot}{3 \left(3 c_{\theta
   }+1\right)}=\mathcal{Q} (1+q-\mathcal{Q})-\frac{r \left(c_a+1\right) \Udot}{3 \left(3 c_{\theta }+1\right)}-\frac{2 \left(c_a+1\right) \mathcal{A} \Udot}{3 \left(3 c_{\theta }+1\right)}+\nonumber \\
	& +\frac{\left(c_a+1\right) \Udot^2}{3 \left(3 c_{\theta
   }+1\right)}+\frac{2 \left(2 c_{\sigma }-1\right) (1-\mathcal{Q})^2}{3 c_{\theta }+1}+\frac{\Omega \left((\gamma -2) v^2-3 \gamma
   +2\right)}{2 \left(3 c_{\theta }+1\right) \left(1-v^2\right)},\label{50.d}\\
& \pmb{\partial}_0 \mathcal{A}  = (q + 3\Sigma)\mathcal{A} - \Udot + r,\\
& \pmb{\partial}_0 \Omega+\frac{(\gamma -2) v\pmb{\partial}_1\Omega}{G_-}-\frac{\gamma  \Omega \pmb{\partial}_1 v}{G_-}= -\frac{2 \gamma  \mathcal{A} v \Omega}{G_-} +\frac{2 q \Omega \left(1-(\gamma -1) v^2\right)}{G_-}+\nonumber \\ 
&-\frac{3 \gamma  \mathcal{Q}
   \left(1-v^2\right) \Omega}{G_-}+\frac{2 (\gamma -2) r v \Omega}{G_-}+\frac{2 \Omega
   \left((1-2 \gamma ) v^2+1\right)}{G_-}\label{43.g},	
\\
&\pmb{\partial}_0 v-\frac{(\gamma -1) \left(1-v^2\right)^2 \pmb{\partial}_1 \Omega}{\gamma 
  \Omega G_- }+\frac{(\gamma -2) v \pmb{\partial}_1 v}{G_-}= \frac{2 (\gamma -1) \mathcal{A} \left(1-v^2\right) v^2}{G_-}+\nonumber \\ & -\frac{3 (\gamma -2) \mathcal{Q} \left(1-v^2\right) v}{G_-}-\frac{2
   (\gamma -1) r \left(1-v^2\right)^2}{\gamma  G_-}+\frac{2 \left(1-v^2\right) v}{G_-}+\Udot \left(1-v^2\right),\label{43.h}\\
&\pmb{\partial}_0 r=\Udot(1+q)+r(2-3\mathcal{Q})+\pmb{\partial}_1 q,
\end{align}
\end{subequations} where 
\begin{align}
\label{(51)}
&q= -\frac{3 \mathcal{K}}{2(2 c_{\sigma }-1)}+\frac{1}{3} \mathcal{A}\Udot \left(\frac{2 \left(c_a+1\right)}{3 c_{\theta }+1}+\frac{1-2 c_a}{1-2
   c_{\sigma }}\right)-\frac{\left(c_a+1\right) \left(3 c_{\theta }+2 c_{\sigma }\right) \pmb{\partial}_1 \Udot}{3 \left(3 c_{\theta }+1\right) \left(2
   c_{\sigma }-1\right)}+\nonumber \\ & +\frac{1}{3} r \left(c_a+1\right) \left(\frac{1}{3 c_{\theta }+1}+\frac{1}{2 c_{\sigma }-1}\right) \Udot+\frac{1}{6}
   \Udot^2 \left(\frac{5 c_a+2}{1-2 c_{\sigma }}-\frac{2 \left(c_a+1\right)}{3 c_{\theta }+1}\right)+\nonumber\\
	& +\frac{\mathcal{A}^2}{2(2 c_{\sigma }-1)}+\frac{4\left(3
   c_{\theta }+2 c_{\sigma }\right) \mathcal{Q}}{3 c_{\theta }+1}+\left(\frac{6-3 c_{\theta }-14 c_{\sigma }}{4 c_{\sigma }-2}+\frac{2-4 c_{\sigma }}{3
   c_{\theta }+1}+1\right) \mathcal{Q}^2+\nonumber \\ 
	& -\frac{\Omega \left(v^2 \left(3 (\gamma +1) c_{\theta }+2 (\gamma -2) c_{\sigma }+3\right)-6
   \gamma  c_{\sigma }-3 c_{\theta }+4 c_{\sigma }+3 \gamma -3\right)}{2 \left(3 c_{\theta }+1\right) \left(2 c_{\sigma }-1\right)
   \left(1-v^2\right)}+\nonumber \\ & +\frac{1-9 c_{\theta }-8 c_{\sigma }}{6 c_{\theta }+2},
\end{align}
subject to the constraints
\begin{subequations}
\label{reduced_syst_2:constr}
\begin{align}
& \pmb{\partial}_1 \mathcal{N}^{-1} = (r - \dot{U}) \mathcal{N}^{-1}, \\
& \pmb{\partial}_1 \mathcal{K} = 2(r + \mathcal{A})  \mathcal{K}, \\
& \pmb{\partial}_1 \mathcal{A}+ c_a \pmb{\partial}_1 \Udot= -\frac{3}{2} \mathcal{K}+r \left(c_a \Udot+\mathcal{A}\right)+2 c_a \mathcal{A}\Udot-\frac{1}{2} c_a
   \Udot^2+\frac{3}{2} \mathcal{A}^2+\nonumber \\ & -\frac{3}{2}\left(3 c_{\theta }+2 c_{\sigma }\right) \mathcal{Q}^2+3\left(2 c_{\sigma }-1\right) \mathcal{Q}+\frac{3 \Omega \left((\gamma -1) v^2+1\right)}{2 \left(1-v^2\right)}-3 c_{\sigma }+\frac{3}{2},\label{52.c}\\
&\pmb{\partial}_1\mathcal{Q}= -\frac{3 \left(2 c_{\sigma }-1\right) \mathcal{A} (1-\mathcal{Q})}{3 c_{\theta }+2 c_{\sigma }}+r \left(\mathcal{Q}+\frac{1-2 c_{\sigma }}{3
   c_{\theta }+2 c_{\sigma }}\right)-\frac{3 \gamma  v \Omega}{2 \left(3 c_{\theta }+2 c_{\sigma }\right) \left(1-v^2\right)}\label{52.d}.	
\end{align}
\end{subequations}


We can choose to study the system \eqref{(43)}, \eqref{(44)}, \eqref{q_r_defs} or  the system \eqref{reduced_syst_2:evol}, \eqref{(51)}, \eqref{reduced_syst_2:constr}, depending on the particular application.

In the comoving aether temporal gauge, which implies a preferred foliation,
the only remaining freedom is the coordinate rescalings $t \rightarrow f(t)$ and 
$x \rightarrow g(x)$ (time and space reparameterization freedom), consistent with
\begin{align}
& \pmb{\partial}_0 E_1^1  = (q + 3\Sigma) E_{1}^1, \\
& \pmb{\partial}_1 \mathcal{N}^{-1} = (r - \dot{U}) \mathcal{N}^{-1},
\end{align}
where we recall that
$ \pmb{\partial}_0 :=  \mathcal{N}^{-1} \partial_t$,\quad
$ \pmb{\partial}_1 := E_{1}^1 \partial_x$.

How do we best treat $\mathcal{N}^{-1}$ in the evolutions eqns? 
There is no evolution eqn for $\mathcal{N}^{-1}$ but, in principle,
we can integrate the                        
spatial constraint and use the time reparameterization to determine
$\mathcal{N}^{-1}$. [This is what happens in GR    
in the separable gauge, where neither an algebraic equation or an                         
evolution equation for $\mathcal{N}^{-1}$ is available; the conditions are sufficient                          
to integrate the spatial derivative constraint equation, whence a time redefinition                 
can be employed to set  $\mathcal{N}^{-1}=1$.]                   
                                               
Depending on the application, we could do one of the following: (i) Since $\mathcal{N}^{-1}$ is positive-definite, we can determine the
qualitative behaviour of the system by simply studying the right-hand-sides of the 
evolution equations. (ii) In many special cases of interest we can integrate  for $\mathcal{N}^{-1}$ 
and replace the left-hand-sides by partial time derivatives (as  in GR    
in the separable gauge). (iii) Numerically we could solve for $\mathcal{N}^{-1}$ in the integration                     
(although this looks messy). (iv) Analytically, we could, for example,           
use the commutators to obtain evolution and constraint equations for $r$, and then ${\mathcal{N}^{-1}            
= \exp(\int (\dot{U} - r)dx)}$ and define new variables  (e.g., use $\dot{U} - r$ as a variable).                                   
(v) While the comoving aether gauge choice is motivated physically,             
it is not ideal for doing analysis and numerics; we could change to a separable gauge.

In practice, it is often useful to choose a gauge in order to compare with the FLRW 
model as easily as possible (e.g., so we can choose integration functions 
for $\mathcal{N}, E_{1}^1$ in the above to be trivial).


As an example, let us consider  McVittie-like models \cite{McVittie}.
The line element is given by 
\begin{equation}
-\left(\frac{1-\frac{M}{2 A x}}{1+\frac{M}{2 A x}} \right)^2 d t^2+A^2\left(1+\frac{M}{2 A x}\right)^4 \left[d x^2 + x^2 (d\y^2 + \sin^2 \y  d\z^2)\right],
\end{equation}
where $A=A(t)$ and $M$ is a constant. 
We recall that as $x\rightarrow +\infty$ the metric approaches the flat FLRW solution and for 
constant $A$ we obtain the Schwarzschild solution. Several aspects of the McVittie
solution, of geometrical and physical relevance have been investigated by many authors \cite{McVittieI}. In GR the McVittie
solution is unique under the following assumptions: (i) The metric is spherically symmetric with a singularity at the centre.
(ii) The matter distribution is a perfect fluid.
(iii) The metric must asymptotically tend to an isotropic cosmological form.
(iv) The fluid flow is shear-free. 
McVittie-like models were investigated, e.g.,  in the references \cite{McVittieII}. 

For this metric we obtain in our scenario: 
\begin{subequations}
\begin{align}
&\udot=-\frac{16 M x^2 A^2}{(M-2 x A) (M+2 x A)^3},\\
& a=\frac{4 x A (M-2 x A)}{(M+2 x A)^3},\\
& K= \frac{16 x^2 A^2}{(2 x A+M)^4},\\
& \R=0,\\
& \S=\frac{64 M x^3 A^3}{(2 x A+M)^6},\\
&\sigma_+=0,\\
&\theta=\frac{3 \dot{A}}{A},
\end{align}
\end{subequations}
which implies that
\begin{align*}
&\mathcal{Q}=1,\Sigma=0, r=0, \\
&q=\frac{A \ddot{A} (2 x A+M)}{\dot{A}^2 (M-2 x A)}-\frac{2 M}{M-2 x
   A}
\end{align*}
For this metric, the equations \eqref{50.c} and \eqref{52.d} lead to a contradiction, unless $c_a=0$ and either $v=0$ or $\Omega=0,$ or both. 
Assuming $c_a=v=\Omega=0$ and substituting into equation \eqref{50.d} we obtain 
\begin{equation}
\frac{A \ddot{A} (2 x A+M)-2 M \dot{A}^2}{\dot{A} (M-2 x A)}=0
\end{equation} 
which, in general, is not satisfied as we vary $t$ and $x$. Thus, assuming $c_a=0$ and $v=0, \Omega\neq 0,$ we obtain 
\begin{equation}
\label{Eq58}
\Omega =\frac{2 \left(3 c_{\theta}+1\right) \left(A \ddot{A} (2 x
   A+M)-2 M \dot{A}^2\right)}{(3 \gamma -2) \dot{A}^2 (M-2 x A)}.
\end{equation}

The equations \eqref{reduced_syst_2:evol} and the constraints \eqref{reduced_syst_2:constr}, with the exception of \eqref{43.g}, \eqref{43.h} and \eqref{52.c}, are identically satisfied. The equations \eqref{43.g} and \eqref{43.h} reduce to 
\begin{subequations}
\begin{align}
&\pmb{\partial}_0 \Omega= (2 q+2-3\gamma)\Omega,\\
& (1-\gamma)\pmb{\partial}_1 \Omega=\gamma \Omega \Udot.
\end{align}
\end{subequations}
which, after the substitution of \eqref{Eq58}, lead to 
\begin{subequations}
\begin{align}
&\ddot{A}= \frac{\dot{A}^2 ((\gamma +1) M-2 (\gamma -1) x A)}{A
   (2 x A+M)},\\
&	\dddot{A}= \frac{\dot{A}^3 \left(4 x A \left((\gamma
   -1) (3 \gamma -1) x A-3 \gamma ^2 M+M\right)+(\gamma  (3 \gamma
   +2)+1) M^2\right)}{A^2 (2 x A+M)^2}.
	\end{align}
\end{subequations}
and the restriction \eqref{52.c} becomes
\begin{equation}
\frac{3 \left(3 c_{\theta}+1\right) \left(\dot{A}^2 ((4-6 \gamma ) x
   A+(3 \gamma +2) M)-2 A \ddot{A} (2 x A+M)\right)}{2 (3 \gamma
   -2) \dot{A}^2 (M-2 x A)}=0.
\end{equation}
The last three equations, in general, are not satisfied simultaneously for all $t$ and $x$ unless we set $\gamma=0,$ which implies
\begin{equation}
A(t)=c_2 e^{c_1 t}, \quad \Omega=1+3 c_\theta.
\end{equation}

Summarizing:
the McVittie-like models only  exist  for the choice of parameters $c_a=0,\gamma=0$, and for aligned aether ($v=0$). Since $\gamma=0$, the matter fluid corresponds to a cosmological constant. 
The solution is characterized by 
\begin{subequations}
\begin{align}
&\udot=-\frac{16 c_2^2 M x^2 e^{2 c_1 t}}{\left(M-2 c_2 x e^{c_1 t}\right)
   \left(2 c_2 x e^{c_1 t}+M\right){}^3},\\
& a=\frac{4 c_2 x e^{c_1 t} \left(M-2 c_2 x e^{c_1 t}\right)}{\left(2 c_2 x
   e^{c_1 t}+M\right){}^3},\\
& K= \frac{16 c_2^2 x^2 e^{2 c_1 t}}{\left(2 c_2 x e^{c_1 t}+M\right){}^4},\\
& \R=0,\\
& \S=\frac{64 c_2^3 M x^3 e^{3 c_1 t}}{\left(2 c_2 x e^{c_1 t}+M\right){}^6},\\
&\sigma_+=0,\\
&\theta=3 c_1.
\end{align}
\end{subequations}
and
\begin{align}
&\mathcal{Q}=1,\Sigma=0, r=0, v=0,  \Omega=1+3 c_\theta, q=-1.
\end{align}

Irrespective of the sign of the expansion ($\theta>0$ or $\theta<0$), the quantities $\udot, a, K, \S$ tend to zero as $t\rightarrow +\infty$.

\newpage

\subsection{Special case:  $\udot=0$}

The evolution equations above were derived under the assumption that $\udot \neq 0$.
We now consider the special case of $\udot=0$ and display the appropriate eqns:

\begin{subequations}
\begin{align}
& \e_0 (\ex) = - \tfrac13 (\theta- 6\sigma_+) \ex,\label{udoteqns1}    
\\
& \e_0 (K) = - \tfrac23 (\theta + 3 \sigma_+)K,
\\
&\e_0(\theta)=-\frac{1}{3} \theta^2+\frac{6(2 c_\sigma-1)\sigma_+^2}{3 c_\theta+1}+\frac{\left(2-3\gamma+(\gamma-2)v^2\right)\hat{\mu}}{2(3 c_\theta+1)(1-v^2)},\\
&\e_0(\sigma_+)=\frac{1}{2}\sigma_+^2-\theta\sigma_+ +\frac{(3 c_\theta+1)\theta^2}{18(2 c_\sigma-1)} -\frac{a^2}{2(2 c_\sigma-1)} +\frac{K}{2(2 c_\sigma-1)}+\nonumber \\ & +\frac{\left((1+\gamma)v^2-1\right)\hat{\mu}}{6(2c_\sigma-1)(1-v^2)},\\
&\e_0(a)=-\frac{1}{3}a (\theta +3 \sigma_+)+\frac{\gamma v \hat{\mu}}{2(1-v^2)},\\ 
&\e_0(\hat{\mu}) -\frac{\e_1
  \left( \hat{\mu}   \right) v   \left(
 2-\gamma\right) }{G_-}-
 \frac {\gamma \hat{\mu}  \e_1 \left( v \right) }{G_-}  = -\frac {2\gamma
 \hat{\mu} v ^{2}
 \sigma_+  }{G_-} +\frac {
 \gamma  \left(v^{2}  -3\right)
 \hat{\mu}  \theta }{3 G_-}-2 \frac{\gamma \hat{\mu}  v a  }{G_-},
 \\  
&   \e_0(v) -{\frac {{
 \e_1} \left( \hat{\mu}   \right)  \left(1-v^2\right) ^{2}
  \left( \gamma-1 \right) }{ \gamma \hat{\mu} G_-  }}-
 {\frac {v   \left(2- \gamma\right) \e_1 \left(
 v   \right) }{G_-}} = {\frac {2 v   \left(1-v^2
  \right)  \sigma_+ }{G_-}}+ \nonumber \\ &{\frac {v
  \left(1-v^2\right)  \left( 3 \gamma-4 \right) \theta  }{ 3 G_-}}   +
 \frac {2 v ^{2} \left(1-v^2\right)  \left(
 \gamma-1 \right) a  }{G_-}.\label{udoteqns1b}
\end{align}
\end{subequations}
Constraints ($3c_\theta+2c_\sigma \neq 0$):
\begin{subequations}
\begin{align}
& \e_1(\ln K) = 2a,\label{udoteqns2}
\\
& \e_1(a)=\frac{G_+ \hat{\mu}}{2    (1-v^2)}-\frac{1}{6}\left(3
   c_{\theta }+1\right) \theta^2-\frac{3}{2}\left(2 c_{\sigma }-1\right) \sigma_{+}^2 -\frac{K}{2}+\frac{3 a^2}{2}
	\\
	& \e_1(\theta)=-\frac{3\gamma c_\sigma v \hat{\mu}}{(3 c_\theta+2 c_\sigma)(1-v^2)},
\\
& \e_1(\sigma_+)=3 a \sigma_+ -\frac{3 \gamma c_\theta v \hat{\mu}}{2(3 c_\theta+2 c_\sigma)(1-v^2)}.\label{udoteqns2b}
\end{align}
\end{subequations}
\subsubsection{Normalized variables} 

The $\beta$-normalized equations are: {\footnote{Strictly speaking, we assume $v \neq 0$ here, since in the dust case below   $v = 0$ 
leads to a contradiction when $\dot{u}=0$. We shall study an exceptional case later. }}
\begin{subequations}
\begin{align}
& \pmb{\partial}_0 E_1^1  = (q + 3\Sigma) E_{1}^1, \\
&  \pmb{\partial}_0 \mathcal{K}  = 2q \mathcal{K}, \\
&  \pmb{\partial}_0 \mathcal{Q}=\mathcal{Q} (1+q-\mathcal{Q})+\frac{2 \left(2 c_{\sigma }-1\right) \Sigma^2}{3 c_{\theta }+1}+\frac{\Omega \left((\gamma -2) v^2-3 \gamma
   +2\right)}{2 \left(3 c_{\theta }+1\right) \left(1-v^2\right)}, \label{eq138}\\
&  \pmb{\partial}_0 \Sigma =\Sigma (1+q-3 \mathcal{Q})+\frac{1}{2} \Sigma^2+\frac{3 \mathcal{K}}{2(2 c_{\sigma }-1)}-\frac{\mathcal{A}^2}{2(2 c_{\sigma
   }-1)}+\frac{\left(3 c_{\theta }+1\right) \mathcal{Q}^2}{2(2 c_{\sigma }-1)}+	\nonumber \\ & +\frac{\Omega \left((\gamma +1) v^2-1\right)}{2 \left(2 c_{\sigma
   }-1\right) \left(1-v^2\right)}, \label{eq139}\\
	& \pmb{\partial}_0 \mathcal{A}=q \mathcal{A}+\frac{3 \gamma  v \Omega}{2\left(1-v^2\right)},\\
& \pmb{\partial}_0 \Omega+\frac{(\gamma -2) v\pmb{\partial}_1\Omega}{G_-}-\frac{\gamma  \Omega \pmb{\partial}_1 v}{G_-}= -\frac{2 \gamma  \mathcal{A} v \Omega}{G_-} +\frac{2 q \Omega \left(1-(\gamma -1) v^2\right)}{G_-}+\nonumber \\ 
&-\frac{3 \gamma  \mathcal{Q}
   \left(1-v^2\right) \Omega}{G_-}+\frac{2 (\gamma -2) r v \Omega}{G_-}+\frac{2 \Omega
   \left((1-2 \gamma ) v^2+1\right)}{G_-},	
\\
&\pmb{\partial}_0 v-\frac{(\gamma -1) \left(1-v^2\right)^2 \pmb{\partial}_1 \Omega}{\gamma 
  \Omega G_- }+\frac{(\gamma -2) v \pmb{\partial}_1 v}{G_-}= \frac{2 (\gamma -1) \mathcal{A} \left(1-v^2\right) v^2}{G_-}+\nonumber \\ & -\frac{3 (\gamma -2) \mathcal{Q} \left(1-v^2\right) v}{G_-}-\frac{2
   (\gamma -1) r \left(1-v^2\right)^2}{\gamma  G_-}+\frac{2 \left(1-v^2\right) v}{G_-},
\end{align}
\end{subequations}
subject to the restrictions: 
\begin{subequations}
\begin{align}
& \pmb{\partial}_1 \mathcal{N}^{-1} = r \mathcal{N}^{-1}, \\
& \pmb{\partial}_1 \mathcal{K} = 2(r + \mathcal{A})  \mathcal{K}, \\
& \pmb{\partial}_1 \mathcal{A}= -\frac{3}{2} \mathcal{K}+r \mathcal{A}+\frac{3}{2} \mathcal{A}^2 -\frac{3}{2}\left(3 c_{\theta }+2 c_{\sigma }\right) \mathcal{Q}^2+3\left(2 c_{\sigma }-1\right) \mathcal{Q}\nonumber \\ & +\frac{3 \Omega \left((\gamma -1) v^2+1\right)}{2 \left(1-v^2\right)}-3 c_{\sigma }+\frac{3}{2},\\
& \pmb{\partial}_1 \mathcal{Q}=r \mathcal{Q}-\frac{3 \gamma  c_{\sigma } v \Omega}{\left(3 c_{\theta }+2 c_{\sigma }\right)
   \left(1-v^2\right)},\label{eq134}\\
& \pmb{\partial}_1 \Sigma=r \Sigma +3 \mathcal{A} \Sigma-\frac{9 \gamma c_\theta  v \Omega}{2 \left(3 c_\theta+2 c_\sigma\right) \left(1-v^2\right)}. \label{eq135} 
\end{align}
\end{subequations}

Using the identity $\Sigma=1-\mathcal{Q},$ and
combining equations \eqref{eq138} and \eqref{eq139} and equations \eqref{eq134} and \eqref{eq135}, respectively, we obtain:
\begin{align}
& q= -\frac{3 \mathcal{K}}{2(2c_{\sigma }-1)}+\frac{\mathcal{A}^2}{2(2 c_{\sigma }-1)}-\frac{\left(3c_\theta +2 c_{\sigma }\right) \left(3c_\theta +8
   c_{\sigma }-3\right) \mathcal{Q}^2}{2 \left(3c_\theta +1\right) \left(2 c_{\sigma }-1\right)} +\frac{4\left(3c_\theta +2 c_{\sigma }\right) 
   \mathcal{Q}}{3c_\theta +1}+\nonumber \\ & -\frac{\Omega \left(v^2 \left(3 (\gamma +1)c_\theta +2 (\gamma -2) c_{\sigma }+3\right)
   -6 \gamma  c_{\sigma }-3c_\theta
   +4 c_{\sigma }+3 \gamma -3\right)}{2 \left(3c_\theta +1\right) \left(2 c_{\sigma }-1\right) \left(1-v^2\right)}+\frac{1-9 c_{\theta
   }-8 c_{\sigma }}{2(3c_\theta +1)}, \label{udot_zero_def_q}
\end{align}
\begin{align}
r=- 3 \mathcal{A} (1- \mathcal{Q})+\frac{3 \gamma  v \Omega}{2(1-v^2)}. \label{udot_zero_def_r}
\end{align}

The final equations for the reduced phase space $\left(\mathcal{N},  E_1^1, \mathcal{K}, \mathcal{Q}, \Omega, v\right)^T$ are the
evolution equations and  restrictions displayed above (less the equations  \eqref{eq139},\eqref{eq135} for the frame derivatives of
$\Sigma$),
where $q$ and $r$ are defined by \eqref{udot_zero_def_q} and \eqref{udot_zero_def_r}, 
respectively. (It will be useful to define $\mathcal{D} \equiv \mathcal{A}^2 -3\mathcal{K}$ in some computations).
Again we note that formally the same equations can be obtained from the previous case ($\dot{u} \neq 0$) by setting $\dot{u}=0$; 
but here we have derived the equations properly.


\subsection{The subset $\dot{U}=v=0$}

Let us consider the special subset  $\dot{U}=v=0$.
We also assume   $c_\theta=0$ and $c_\sigma \neq 0$. We first assume that $\gamma \neq 1$.
The final equations for the reduced phase space are then:
\begin{subequations}
\begin{align}
& \pmb{\partial}_0 E_1^1  = (q + 3(1-\mathcal{Q})) E_{1}^1, \\
&  \pmb{\partial}_0 \mathcal{K}  = 2q \mathcal{K}, \\
&  \pmb{\partial}_0 \mathcal{Q}=\mathcal{Q} (1+q-\mathcal{Q})+{2 \left(2 c_{\sigma }-1\right) (1-\mathcal{Q})^2}+\frac{1}{2} \Omega \left(2-3 \gamma
 \right),\\
	& \pmb{\partial}_0 \mathcal{A}=q \mathcal{A},\\
& \pmb{\partial}_0 \Omega= (2 q -3 \gamma  \mathcal{Q} +2)\Omega,
\end{align}
\end{subequations}
subject to the restrictions: 
\begin{subequations}
\begin{align}
& \pmb{\partial}_1 \mathcal{N}^{-1} = r \mathcal{N}^{-1}, \\
& \pmb{\partial}_1 \mathcal{K} = 2(r + \mathcal{A})  \mathcal{K}, \\
& \pmb{\partial}_1 \mathcal{A}= -\frac{3}{2} \mathcal{K}+r \mathcal{A}+\frac{3}{2} \mathcal{A}^2 -3 c_{\sigma }\mathcal{Q}^2
+\frac{3}{2}\left(2 c_{\sigma }-1\right)(2 \mathcal{Q} -1) +\frac{3 }{2}\Omega,\\
& \pmb{\partial}_1 \mathcal{Q}=r \mathcal{Q}, \\
&\pmb{\partial}_1 \Omega=  2r \Omega,
\end{align}
\end{subequations}
where $q$ and $r$ are defined by:

\begin{align}
& q= \frac{1}{2(2c_{\sigma }-1)}\Big\{-3\mathcal{K}+\mathcal{A}^2-2 c_{\sigma }\left(8c_{\sigma }-3\right) \mathcal{Q}^2 
+ 16c_{\sigma }(2c_{\sigma }-1)\mathcal{Q} \nonumber \\ 
& -\Omega \left(-2c_{\sigma }(3 \gamma - 2) +3(\gamma -1)\right)+(1-8 c_{\sigma })(2c_{\sigma }-1)\Big\}.
\end{align}

\begin{align}
r=- 3 \mathcal{A} (1- \mathcal{Q}).
\end{align}
We recall that
$\pmb{\partial}_0 E_1^1  = (q + 3(1-\mathcal{Q})) E_{1}^1,$
$\pmb{\partial}_1 \mathcal{N}^{-1} = r \mathcal{N}^{-1},$
where
$ \pmb{\partial}_0 :=  \mathcal{N}^{-1} \partial_t$,\quad
$ \pmb{\partial}_1 := E_{1}^1 \partial_x$.
[The only remaining freedom is the coordinate rescalings $t \rightarrow f(t)$ and 
$x \rightarrow g(x)$].~\footnote{
$\mathcal{A}^2$ and $\mathcal{K}$ only appear in the evolution equations via the combination  
$\mathcal{D} \equiv \mathcal{A}^2 -3\mathcal{K}$ (but also appear in the constraints; e.g., via $r$).}

\subsubsection{The special case $\gamma = 1$}

In the special case of dust, we lose the equation $\pmb{\partial}_1 \Omega=  2r \Omega$, but all of the remaining equations are valid with 
$\gamma = 1$.

\newpage 
 
 \section{Lema\^{\i}tre-Tolman-Bondi model}

 The Lema\^{\i}tre-Tolman-Bondi (LTB) model in GR
 \cite{lemaitre} is the spherically symmetric dust
 solution of the Einstein equations which can be regarded as a
 generalization of the FLRW universe. LTB metrics with dust source
 and a comoving and geodesic 4-velocity constitute a well known
 class of exact solutions of Einstein's field equations \cite{LTB}. 
 
 The line element
 for a spherically symmetric comoving dust is:
 
 \be\label{LTB}
 ds^2 = - dt^2 + \frac{ [R'(t,x)]^2 }{ 1+2E  } dx^2 + R(t,x)^2 
 [ d\theta^2 + \sin^2\theta d\phi^2 ]
 \ee 
 (where an overdot denotes a $t$-derivative and a prime denotes an $x$-derivative).
 The geometric variables, $\{ H,\ \sp,\ \R,\ \S \}$, are defined by
 \be
 H+\sp = \frac{\dot{R}}{R};~H-2\sp = \frac{\dot{R}'}{R'};~\R = -\frac{4(ER)'}{R^2R'};~\S = -\frac{1}{12} \R - \frac{E}{R^2}.
 \ee  where 
 \begin{equation}
 \R=4\e_1 a -6 a^2+2 K,\; \S=-\frac{1}{3}\e_1 a+\frac{1}{3}K,
 \end{equation}
 \begin{equation}
 a=-\frac{\sqrt{1+E }}{R(t,x)}, \; K=\frac{1}{R(t,x)^2}.
 \end{equation}
 
 The matter variable is defined by $\rho= \frac{2M'}{R^2R'}$.
 The 3 free functions $R(t,r)$, $E $, $M $ are subject to the
 Gauss constraint.
 We can introduce Hubble-normalized variables \cite{WAIN}:
 \be
 \Sp = \frac{\sp}{H},\quad \Omega = \frac{\rho}{3H^2},\quad \Omega_k = 
 - \frac{\R}{6H^2},\quad \calS = \frac{\S}{3H^2},
 \ee
 where (the deceleration parameter) $q = 2 \Sp^2 + \frac12 \Omega$, and $D = \Omega_k - 6 \calS$.
 By introducing a new time variable $\tau(t,r)$ defined by
 ${\partial t}/{\partial \tau} = {1}/{H}$, where $H>0$, $H$ decouples from the evolution equations.
 Using these variables, expressed in the coordinates above,
the evolution equations for GR are (the dynamical system) (see also \cite{Sussman2}):
 \footnote{We note that the exact GR LTB is not a solution to the Einstein-aether equations (see section 4).}
 
 \begin{subequations}
 	\begin{align}
 	\partial_\tau \Omega &= (\Omega + 4\Sp^2 -1)\Omega,
 	\\
 	\partial_\tau \Sp &= (\frac12\Omega + 2\Sp^2 - 2)\Sp - \frac12(1-\Omega-\Sp^2-\D),
 	\\
 	\partial_\tau \D &= (\Omega + 4\Sp^2-2\Sp)\D.
 	\end{align}
 \end{subequations}
 The flat FLRW equilibrium point is given by $(\Omega,\Sp,\D)=(1,0,0)$, where
 $(\Omega_k,\calS)=(0,0)$. It is a saddle with
 eigenvalues  $(1,-\frac32,1)$ \cite{WAIN}.

\newpage

\noindent
{\em Acknowledgements}.
We would like to thank  W. Donnelly, T.~Jacobson and D. Garfinkle, W. C. Lim and C. Uggla for helpful comments. 
A. C. was supported, in part, by NSERC of Canada, and
G.L. was supported by COMISI\'ON NACIONAL DE CIENCIAS Y TECNOLOG\'IA through Proyecto FONDECYT DE POSTDOCTORADO 2014  grant  3140244.


\end{document}